\documentclass[journal]{IEEEtran}

\usepackage{cite}

\long\def\comment#1{}

\newfont{\bbb}{msbm10 scaled 700}

\newfont{\bb}{msbm10 scaled 1100}

\newcommand{\PP}{\mbox{\bb P}}
\newcommand{\RR}{\mbox{\bb R}}

\newcommand{\EE}{\mbox{\bb E}}

\newcommand{\av}{{\bf a}}
\newcommand{\bv}{{\bf b}}

\newcommand{\qv}{{\bf q}}
\newcommand{\rv}{{\bf r}}
\newcommand{\sv}{{\bf s}}
\newcommand{\tv}{{\bf t}}
\newcommand{\uv}{{\bf u}}

\newcommand{\vv}{{\bf v}}
\newcommand{\xv}{{\bf x}}
\newcommand{\yv}{{\bf y}}
\newcommand{\zv}{{\bf z}}

\newcommand{\onev}{{\bf 1}}

\newcommand{\Am}{{\bf A}}
\newcommand{\Bm}{{\bf B}}

\newcommand{\Qm}{{\bf Q}}
\newcommand{\Rm}{{\bf R}}

\newcommand{\Ac}{{\cal A}}

\newcommand{\Dc}{{\cal D}}
\newcommand{\Ec}{{\cal E}}

\newcommand{\Hc}{{\cal H}}
\newcommand{\Ic}{{\cal I}}

\newcommand{\Nc}{{\cal N}}
\newcommand{\Oc}{{\cal O}}

\newcommand{\Sc}{{\cal S}}
\newcommand{\Tc}{{\cal T}}

\newcommand{\Wc}{{\cal W}}

\newcommand{\trace}{{\hbox{tr}}}

\renewcommand{\arg}{{\hbox{arg}}}

\newcommand{\SNR}{{\sf SNR}}

\newcommand{\eqdef}{\stackrel{\Delta}{=}}

\newcommand{\trasp}{{\sf T}}

\usepackage{algorithm,algcompatible}

\usepackage{graphicx,color,epsfig,rotating,subfigure}
\usepackage{amsfonts,amsmath,amssymb}
\usepackage{algorithm}
\usepackage{subfigure}
\usepackage{algpseudocode}

\DeclareMathOperator*{\argmax}{\arg\!\max}% 
\algnewcommand\INPUT{\item[\textbf{Input:}]}%
\algnewcommand\OUTPUT{\item[\textbf{Output:}]}%

\newtheorem{theorem}{Theorem}

\newtheorem{definition}{Definition}

\newtheorem{lemma}{Lemma}

\newtheorem{corollary}{Corollary}
\newcommand{\argmin}{\operatornamewithlimits{argmin}}

\newtheorem{example}{Example}

\title{Greedy Sparse Signal Recovery Algorithm\\Based on Bit-wise MAP detection}

\author{
\IEEEauthorblockN{
              Jeongmin Chae and Song-Nam Hong}\\
\IEEEauthorblockA{Ajou University, Suwon, Korea,\\
              email:\{jmchae92, snhong\}@ajou.ac.kr}
}

\begin{document}
\maketitle

%\date{}

%%%%%%%%%%%%%%%%%%%%%%%%%%%%%%%%%%%%%%%%%%%%%%%%%%%%
\begin{abstract}  
We propose a novel greedy algorithm for the support recovery of a sparse signal from a small number of noisy measurements. In the proposed method, a new support index is identified for each iteration based on {\em bit-wise maximum a posteriori} (B-MAP) detection. This is optimal in the sense of detecting one of the remaining support indices, provided that all the detected indices in the previous iterations are correct. Despite its optimality, it requires an expensive complexity for computing the maximization metric (i.e., a posteriori probability of each remaining support) due to the marginalization of high-dimensional sparse vector. We address this problem by presenting a good proxy (named {\em B-MAP proxy}) on the maximization metric which is accurate enough to find the maximum index, rather than an exact probability, Moreover, it is easily evaluated only using vector correlations as in orthogonal matching pursuit (OMP), but the use completely different proxy matrices for maximization. We demonstrate that the proposed B-MAP detection provides a significant gain compared with the existing methods as OMP and MAP-OMP, having the same complexity. Subsequently, we construct the advanced greedy algorithms, based on B-MAP proxy, by leveraging the idea of 
compressive sampling matching pursuit (CoSaMP) and subspace pursuit (SP). Via simulations, we show that the proposed method outperforms also OMP and MAP-OMP under the frameworks of the advanced greedy algorithms. %Thus, B-MAP detection would be a better alternative of the counterparts OMP and MAP-OMP for the support recovery of a sparse signal.
\end{abstract}

\vspace{0.1cm}

%%%%%%%%%%%%%%%%%%%%%%
\section{Introduction}\label{sec:intro}

An inverse problem is widely studied in which a vector signal  $\xv \in \RR^N$ is recovered from a set of linear noisy measurements 
\begin{equation}
\yv = \Am\xv + \zv,
\end{equation} with an $M \times N$ measurement matrix $\Am$. In particular when $M < N$ (i.e., underdetermined system), this problem has infinite solutions and thus, it can be solved only when some additional a priori information of $\xv$ is available. 
In \cite{Candes 2006,Donoho2006_2}, it was proved that $\xv$ can be exactly reconstructed with a priori knowledge on the sparsity of 
 $\xv$ (i.e., $\|\xv\|_0 = K$ with $K \ll N$), where $K$ is called sparsity-level. Also, the optimal sparse signal can be obtained by solving $\ell_0$-minimization  as
\begin{equation}
    \xv^{\star}=\arg\min_{\xv} \|\xv\|_0 \indent \text{ subject to } \quad \|\yv-\Am\xv\|_2\leq \eta, \label{eq:zeronormformulation}
\end{equation}where $\|{\bf x}\|_0$ is introduced to ensure the sparsity of $\xv$. In general, the $\ell_0$-minimization is known to be NP-hard  \cite{Garey}. Leveraging the idea of convex optimization, a well-established method, called least absolute shrinkage and selection operator (LASSO) (a.k.a., basis pursuit denoising (BPDN)), was proposed in \cite{Candes2007,Candes2008,Chen1995,Chen1999,Tibshirani}, where $\ell_1$-norm is used as a convex-relaxation of $\ell_0$-norm. LASSO is usually solved via convex-optimization techniques such as iterative shrinkage thresholding method (ISTA), alternating direction method of multipliers (ADMN), coordinate descent (CD), and primal-dual interior-point method (PDIPM). They can provide the stability and uniform guarantees but have polynomial complexities. 

%%%%%%%%%%%%%%%%%%%%%%%%%%%%
%%%%%%%% GREEDY APPROACH %%%%%%%%
%%%%%%%%%%%%%%%%%%%%%%%%%%%%

A greedy approach seems to be attractive due to its lower complexities than convex-based algorithms in sparse signal recovery problems. The underlying idea of greedy-based algorithms is to estimate the support of a sparse signal in a sequential fashion. Namely, a new index is added to a target support for each iteration by solving the so-called sub-optimization problem. Since it has much lower complexity than the overall vector-wise optimization, the greedy approach can significantly reduce the computational complexity. Orthogonal matching pursuit (OMP) is the most popular greedy method in which the sub-optimization problem is just to find an index having the maximum correlation with a residual vector  \cite{Tropp2007, Davenport2010, Cai2011, Zhang2011}. This metric in the maximization is referred to {\em OMP proxy}. Furthermore, to overcome the inherent drawback of OMP, several advanced greedy algorithms have been proposed as  stagewise orthogonal matching pursuit (StOMP) \cite{Donoho2012}, iterative hard thresholding (IHT) \cite{Blumensath2009}, compressive sampling matching pursuit (CoSaMP) \cite{Needell2010}, subspace pursuit (SP) \cite{Dai2009} and generalized OMP \cite{Wang2012}. The main idea lies in the selection of multiple support indices (based on OMP proxy) for each iteration, thus being able to degrade the miss-detection probability. However, these approaches may not be optimal due to the use of OMP proxy, if a probabilistic model for a sparse signal recovery is provided.

%%%%%%%%%%%%%%%%%%%%
% PROBABILISTIC APPROACH %%%
%%%%%%%%%%%%%%%%%%%%

Bayesian matching pursuit (BMP) was proposed in \cite{Ji2008,Schniter2008,Zayyani2009,Herzet2010,Dremeau2011} where it exploits 
the probabilistic knowledge on a sparse signal and noise for a support detection. In \cite{Schniter2008}, fast BMP (FBMP) estimates a sparse signal using probabilistic model selections and the associated parameter estimations. Also in \cite{Herzet2010,Dremeau2011}, a sparse signal is updated for each iteration so that the enhancement of a local likelihood function is maximized by assuming a Bernoulli-Gaussian sparse signal models. These methods can show a better recovery probability than the matching pursuit algorithms but they are restricted to certain distributions of a sparse signal. Recently, MAP-OMP was proposed in  \cite{LEE2008}, where OMP proxy is improved using the probabilistic modes of a measurement matrix and a sparse signal. The corresponding proxy is called {\em MAP-ratio proxy}, it relies on the log-MAP ratios of OMP proxy.  It was demonstrated in \cite{LEE2008} that MAP-OMP can outperform OMP with the same computational complexity, provided that a Gaussian measurement matrix is used. 

%%%%%%%%%%%%%%%%%%%%
%  OUR CONTRIBUTIONS
%%%%%%%%%%%%%%%%%%%%

In this paper, we propose a novel greedy algorithm for the support recovery of a sparse signal. The underlying idea of the proposed method is to determine  a new support index for each iteration, based on {\em bit-wise maximum a posteriori} (B-MAP) detection. This is {\em optimal} in the sense of detecting one of the remaining support indices, provided that all the detected ones in the previous iterations are correct. In other words, this detection is optimal under the greedy framework, thus being able to outperform the existing counterparts as OMP and MAP-OMP. The challenging part of B-MAP detection is to compute a posteriori probability for each candidate support index (in short, maximization metric) given the probability distributions of a sparse signal and an additive noise, since it requires a heavy marginalization of high-dimensional sparse vector. Our contribution is to present a good proxy (named {\em B-MAP proxy}) of this metric and verify that it is accurate enough to find the maximum one, rather than an exact probability. Furthermore, B-MAP proxy has the same complexity with the popular OMP proxy, as both only need vector correlations. With this proxy, it is shown that the proposed greedy algorithm can exactly recover a $K$-sparse binary signal within $K$ iterations almost surely, provided that the number of measurements scales with
\begin{equation}
M=\Oc((1+1/\SNR)K\ln(N)),
\end{equation}where $\SNR = \EE\|\xv\|^2/\EE[\|\zv\|^2$ denotes the (vector-wise) signal-to-noise ratio. This achieves the almost same scaling with MAP-OMP in \cite{LEE2008}, both extending the existing statistical guarantees in \cite{Tropp2007} by including the impact of noise. Beyond the asymptotic analysis, we demonstrate that for practical settings (i.e., finite $M$ and $N$), the proposed algorithm yields a significant gain compared with MAP-OMP as well as OMP, having the same computational complexity. Also, the proposed method can ensure a good recovery performance for various types of measurement matrices as in OMP, while MAP-OMP is restricted to Gaussian measurement matrices. We then propose the advanced greedy algorithms, called  B-CoSaMP and B-SP, by replacing OMP proxy with the proposed B-MAP proxy based on CoSaMP \cite{Needell2010} and SP \cite{Dai2009}, respectively. Simulation results demonstrated that the proposed iterative algorithms outperform the corresponding counterparts based on OMP proxy and MAP-ratio proxy.

%%%%%%%%%%%%%%%%%%%
The remainder of this paper is organized as follows. In Section~\ref{sec:Preliminaries}, we provide some useful notations, define our sparse recovery problem, and propose a greedy framework based on B-MAP detection. In Section~\ref{sec:B-MAP}, we derive a good proxy (named B-MAP proxy) which is simply evaluated with the same complexity of OMP proxy. Section~\ref{section:AAE} performs an asymptotic analysis of the proposed greedy algorithm, yielding the scaling-law of the required measurements to ensure a perfect recovery.  In Section~\ref{sec:extension}, advanced greedy algorithms, based on B-MAP proxy, are proposed, which are named B-CoSaMP and B-SP. 
Simulation results are provided in Section~\ref{sec:simulations} to verify the superiorities of the proposed greedy algorithms. Section~\ref{sec:conclusion} concludes the paper.

%%%%%%%%%%Problem Statement%%%%%%%%%%%
\section{Preliminaries}\label{sec:Preliminaries}

In this section, we provide some useful notations which will be used throughout the paper, and describe a sparse signal recovery problem. Then, we define the proposed greedy framework based on bit-wise MAP (B-MAP) detection.

%In this section, we provide some useful notations and define the sparse signal recovery problem, which will be used throughout the paper.

\subsection{Notations}

Let $[N]\eqdef\{1,...,N\}$ for any positive integer $N$. We use $\xv$ and $\Am$ to denote a column vector and matrix, respectively. Also, for a vector $\xv \in \RR^N$, $x_i$ denotes the $i$-th component of $\xv$ for $i \in [N]$. Likewise, for a matrix $\Bm\in\RR^{M\times N}$, the $(i,j)$-th component of $\Bm$ is denoted as $\Bm_{i,j}$. Also, the $i$-th column and row vectors of $\Bm$ are denoted by $\Bm(i,:)$ and $\Bm(:,i)$, respectively. For any positive integer $K \leq N$, the set of all length-$N$ binary vectors with the sparsity-level $K$ is defined as $\Omega$, namely,
\begin{equation}
\Omega \eqdef \left\{\xv\in \{0,1\}^N: \|\xv\|_0 = K\right\}.
\end{equation} 
Also, given an index subset $\Ic \subseteq [N]$, the subset of $\Omega$ is defined as
\begin{equation}
\Omega_\Ic \eqdef \{\xv\in \{0,1\}^N: \|\xv\|_0 = K, x_{i} = 1 \mbox{ for } i\in \Ic\},\label{eq:omega_sub}
\end{equation} where the size of the subset is equal to $|\Omega_\Ic| = {(N-|\Ic|) \choose (K-|\Ic|)}$. For a vector $\xv\in\RR^N$, $\Sc(\xv)$ represents its support which contains the indices of non-zero components of $\xv$, i.e., 
\begin{equation}
\Sc(\xv) \eqdef \{i|x_i \neq 0, i\in [N]\}.
\end{equation} 
As an extension to the set of vectors, we also define the 
\begin{equation}
\Sc(\Omega_{\Ic}) \eqdef \{\Sc(\xv)|\xv \in \Omega_{\Ic}\}.
\end{equation} 
Given any two subsets $\Sc$ and $\Tc$, the difference between two sets $\Sc$ and $\Tc$ are denoted by $\Sc\setminus\Tc$. Also, for a given index subset $\Tc$, we use the subscript notations $\xv_{\Tc}$ (resp. $\xv_{|\Tc}$) and $\Am_{\Tc}$ (resp. $\Am_{|\Tc}$) to represent the subvector of $\xv$ and column submatrix of $\Am$, respectively, which contain only the components and columns whose indices are belong (resp. not belong) to $\Tc$, respectively.
Finally, we let $\onev$ denote the all ones vector.

\subsection{Problem Formulation}

We consider a $N$-dimensional sparse signal recovery problem from a noisy measurement.  Let $\xv\in\RR^N$ be a $K$-sparse signal (i.e., $\|\xv\|_0 = K$). Then, a noisy measurement vector $\yv \in \RR^M$ is obtained as
\begin{equation}
\yv = \Am \xv + \zv, \label{eq:systemmodel}
\end{equation} where $\Am=[\av_1,\av_2,\cdots,\av_N] \in \RR^{M\times N}$ denotes a measurement matrix and $\zv \in \RR^{M}$ denotes the additive noise vector whose components are independent and identically distributed as Gaussian random variables with zero mean and variance $\sigma^2$. For this model, our goal is to recover the support of the sparse signal $\xv$ using a noisy measurement $\yv$ and a measurement matrix $\Am$. The support vector of $\xv$ is denoted as $\sv \in \{0,1\}^N$ with  $s_i = 1$ if $x_i \neq 0$ for $i\in[N]$. Throughout the paper, we assume that the sparsity-level $K$ is known as a priori information and in some cases, the marginal probability mass functions (PMFs) of a support random variable $s_i$ are given as
\begin{equation}
\PP(s_i = 1)\eqdef p_i \in [0,1] \mbox{ for } i\in [N]. \label{eq:apriori_prob}
\end{equation} Note that if it is not mentioned specifically, uniform distribution (i.e., $p_i = 0.5$ for $i \in [N]$) is implicitly assumed. For the case of $s_i = 1$ (i.e., $i \in \Sc(\xv)$), its value $x_i$ is generated by a continuous probability density function (PDF) $f_{x_i}$. We remark that the system model in  (\ref{eq:systemmodel}) is fully described by random variables $\xv$, $\sv$, $\zv$, and $\yv$. Also, the conditional PDF $f_{\yv|\xv}$ is known  as
\begin{equation}
f_{\yv|\xv}(\yv|\xv)  = \prod_{j=1}^{M}f_{y_j|\xv}(y_j|\xv),\label{PDF:channel}
\end{equation} where the equality follows the mutually independence of $z_i$'s given $\xv$, and
\begin{equation}
f_{y_j|\xv}(y_j|\xv)=\frac{1}{\sqrt{2\pi\sigma^2}}\exp\left(-\frac{(y_j - (\Am\xv)_j)^2}{2\sigma^2}\right).\label{eq:s_PDF}
\end{equation} Given the above probabilistic model, we would like to investigate the maximum a posteriori (MAP) support recovery, which is mathematically formulated as
\begin{align}
\hat{\Ic}= \argmax_{\Ic \in \Sc(\Omega)} \;\; \log{\PP(\Sc(\xv) = \Ic\big|\yv,\Am)}. \label{eq:optimal_MAP}
\end{align} 
In general, it is too complicated to solve the above problem due to the combinatorial nature. Specifically, it is required to check the maximization metric in (\ref{eq:optimal_MAP}) for all the $N \choose K$ plausible candidates. This problem will be addressed by presenting a novel greedy framework based on B-MAP detection (see Section~\ref{sec:proposedmethod}).

%%%%%%%%%%%%%%%%%%
%          ALGORITHMS                  %
%%%%%%%%%%%%%%%%%%

\begin{algorithm}
\caption{B-MAP: Fixed non-zero value}
\begin{algorithmic}[1]
\State {\bf Input:} Measurement matrix $\Am \in \RR^{M\times N}$, noisy observation $\yv \in \RR^{M}$, signal value $\beta$, noise variance $\sigma^2$, sparsity-level $K$, and a priori probabilities $\{p_j: j\in [N]\}$.
\vspace{0.05cm}
\State {\bf Output:} Support $\hat{\Ic}^{(K)} = \{\hat{i}_1,...,\hat{i}_K\}$.
\vspace{0.05cm}
\State {\bf Initialization:} $\hat{\Ic}^{(0)} = \phi$ and $\rv^{(1)} = \yv$.% and $\Am_{|\hat{\Ic}^{(0)}} = \Am$.
\vspace{0.05cm}
\State {\bf Iteration:} $k=1:K$
\begin{itemize}
%\item Set $\Qm = \Am_{|\hat{\Ic}^{(k-1)}}$ and $\tv = \rv^{(k-1)}$.
%\item Set $i'$ as the minimum element of  $\hat{\Ic}^{(k-1)}$.
\item Update the B-MAP proxy:
\begin{align*}
\gamma^{(k)}_j &=\log\left(\frac{p_j}{1-p_j}\right)^{(1-\lambda_k)}+ \frac{1}{\sigma^2}\qv^{\trasp}\av_j -\frac{1}{2\sigma^2}\tau\|\av_j\|_2^2.   
\end{align*} where 
\begin{align*}
\qv&= \beta(1-\lambda_k)\rv^{(k)} - \beta^2\lambda_k(1-\lambda_{k+1})\Am_{|\hat{\Ic}^{(k-1)}}\onev \\
\tau&=\beta^2(1- 3\lambda_k + 2\lambda_k\lambda_{k+1}).
%\sum_{i \notin \hat{\Ic}^{(k-1)}} \av_i\\
%\tau =\beta^2(1- 3\lambda_k + 2\lambda_k\lambda_{k+1}).
\end{align*}
\item Select the largest index of B-MAP proxy:
\begin{align*}
       &\hat{i}_k = \argmax \left\{\gamma^{(k)}_j: j \in [N]\setminus \hat{\Ic}^{(k-1)}\right\}.
\end{align*} 
\item Merge the support: $\hat{\Ic}^{(k)} = \hat{\Ic}^{(k-1)} \cup \{\hat{i}_k\}$.
\item Update the residual vector for the next iteration: $\rv^{(k+1)} = \rv^{(k)} -  \beta \av_{\hat{i}_k}$.
\end{itemize}
\end{algorithmic}
\end{algorithm}
%%%%%%%%%%%%%%%%%%%%%%%%%%%%%%%

%%%%%%%%%%%%%%%%%%%%%%%%%%%%%%%%%%%%%%%%%%%%%%%%%
\subsection{The Proposed Greedy Framework} \label{sec:proposedmethod}

We propose a novel greedy algorithm which efficiently solves MAP support recovery problem in  (\ref{eq:optimal_MAP}). The key idea of the proposed method is to perform the factorization of a posteriori probability (i.e., objective function) in 
(\ref{eq:optimal_MAP}) with respect to the support $\Sc(\xv)$. This enables to determine support indices in a greedy manner. Specifically, using the chain rule,  the objective function in (\ref{eq:optimal_MAP}) can be factorized as
\begin{align}
&\log{\PP(\Sc(\xv)= \Ic^{(K)}\big| \Am, \yv )}\nonumber\\
&\;\;\;\;\;\;\;\; =\sum_{k=1}^{K}\log{\PP(i_k\in \Sc(\xv)|\Ic^{(k-1)}\subset\Sc(\xv),\yv,\Am)},\label{eq:obj_fac}
\end{align} where the sequence of index subsets $\Ic^{(k)}$, $k=0,1,...,K$ is defined as $\Ic^{(k)} =\{i_1,...,i_k\}$  for $k=1,...,K$, 
with initial set $\Ic^{(k)} = \phi$. Based on this, the fundamental principle of the proposed greedy algorithm is to determine a new support index at each iteration by taking the solution of 
\begin{equation}
\hat{i}_{k+1} = \argmax_{i \in [N]\setminus \hat{\Ic}^{(k)}}\log{\PP(i\in \Sc(\xv)|\hat{\Ic}^{(k)} \subset \Sc(\xv),\yv,\Am)}, \label{eq:sub_opt}
\end{equation} where $\hat{\Ic}^{(k)}=\{\hat{i}_1,....,\hat{i}_{k}\}$ denotes the support indices chosen during the previous $k$ iterations. The maximization problem in (\ref{eq:sub_opt}) is referred to as B-MAP detection, which is optimal in the sense of detecting one support index from $\Sc(\xv)\setminus\hat{\Ic}^{(k)}$, with the assumption of $\hat{\Ic}^{(k)} \subset \Sc(\xv)$. It is noticeable that the search-space of the proposed greedy algorithm grows linearly with $K$ while that of (vector-wise) MAP detection in (\ref{eq:optimal_MAP}) is exponential. In this way, the search-space complexity of the optimal MAP detection is completely addressed by keeping an optimality in the other sense. Nevertheless, the proposed B-MAP detection still has the computational complexity as the evaluation of the objective function in (\ref{eq:sub_opt}) requires the marginalization of a large-scale dimensional vector (i.e., the summations of all possible $K$-sparse support signals  $\sv \in \Omega_{\hat{\Ic}^{(k)}}$). This problem will be addressed in the following section.

%%%%%%%%%%%%%%%%%%%%%%%%%%%%%%
\section{The Proposed B-MAP Proxy}\label{sec:B-MAP}

In this section we propose a good proxy of the objective function (i.e., bit-wise a posteriori probability) in (\ref{eq:sub_opt}) by assuming a constant-value sparse signal (e.g., binary sparse signal). Namely, it is assumed that  $x_i = \beta\neq0$ with probability 1 for $i \in \Sc(\xv)$. Before stating our main results, we first provide the useful notations and definitions below:
\vspace{0.1cm}
 \begin{definition}\label{def:map}  We define one-to-one mapping to rearrange the indices not belong to $\hat{\Ic}^{(k)}$ as
  \begin{equation}
  \pi_{k}(i):  [N]\setminus \hat{\Ic}^{(k)} \rightarrow [N-k]. \label{eq:map}
 \end{equation} Also, its inverse mapping is defined as $\pi_k^{-1}$. Given two PMFs $p$ and $q$, Kullaback-Leiber (KL) divergence is denoted as $\Dc_{\rm KL}(p||q)$. For $0\leq \epsilon \leq 1$, ${\rm Bern}(\epsilon)$ stands for the Bernoulli distribution with $\PP(x=1)=\epsilon$. Also, to simplify the expressions, we use the notation $\lambda_k$ throughout the paper,  given by
\begin{equation}
\lambda_k \eqdef \frac{K-k}{N-k} \mbox{ for } 1\leq k\leq K. \label{eq:lambda}
\end{equation} Given $\hat{\Ic}^{(k-1)}$, the residual vector for the $k$-th iteration is defined as $\rv^{(k)} = \yv - \sum_{j \in \hat{\Ic}^{(k-1)}} \beta \av_j = \yv - \beta\Am_{ \hat{\Ic}^{(k-1)}}\onev$.\flushright\QED
%\begin{equation}
 %\rv^{(k)} = \yv - \sum_{j \in \hat{\Ic}^{(k-1)}} \beta \av_j = \yv - \beta\Am_{ \hat{\Ic}^{(k-1)}}\onev.
 %\end{equation} \flushright\QED
 \end{definition}
With the above definition, we have:
%%%%%%%%%%%%%%%%%%%%%%%%
%\begin{equation}
%\alpha_j = \begin{cases}
%1, & j \in \Ic\\
%\lambda_k & {\rm else},
%\end{cases}
%\end{equation} and
\vspace{0.1cm}
\begin{theorem}\label{thm:lower} For any $i_k \notin \hat{\Ic}^{(k-1)}$, the lower-bound on the objective function of B-MAP detection is derived as
\begin{align}
&\log{\PP\left(i_k \in \Sc(\xv)|\hat{\Ic}^{(k-1)} \subset \Sc(\xv),\yv,\Am\right)} + C_1 \nonumber\\
&\;\;\geq \sum_{j \in [N]} - \mathcal{D}_{{\rm KL}}\left({\rm Bern}(\alpha_{j})\big|\big|{\rm Bern} (p_{j})\right)\nonumber\\
&\;\;\;\;\; +\frac{\beta}{\sigma^2}\left(\av_{i_k} + \lambda_k \Am_{|\Ic}\onev\right)^{\trasp}\rv^{(k)} -\frac{\beta^2}{2\sigma^2}\trace(\Qm\Rm^{(i_k)}), \label{eq:lower}
%&\eqdef h_{\rm L}\left(i_k|\hat{\Ic}^{(k-1)},\yv,\Am\right), \label{eq:lower}
\end{align} where  $\Ic = \hat{\Ic}^{(k-1)}\cup\{i_k\}$, $\Qm=(\Am_{|\hat{\Ic}^{(k-1)}})^{\trasp}\Am_{|\hat{\Ic}^{(k-1)}}$, the constant $C_1$, which does not depend on $i_k$, is given in (\ref{eq:C1}), $\alpha_j=1$ for $j \in \Ic$ and $\alpha_j=\lambda_k$, otherwise, and the components of $\Rm^{(i_k)}$ are given as
\begin{equation}
\Rm^{(i_k)}_{i,j} =  \begin{cases}
1, & i=j=\pi_{k-1}(i_k)\\
\lambda_k\lambda_{k+1}, & i,j \neq \pi_{k-1}(i_k) \mbox{ and } i\neq j\\
\lambda_k, & \mbox{ else}.
\end{cases} 
\end{equation}
\end{theorem}
\begin{IEEEproof} The proof is provided in Section~\ref{proof:thm1}.
\end{IEEEproof} 
\vspace{0.1cm}
Under the maximization in (\ref{eq:sub_opt}), we only need to consider the terms associated with $i_k$ in (\ref{eq:lower}) as the other constant terms with respect to $i_k$ do not impact on the maximization. Based on this, we can further simplify the lower-bound on (\ref{eq:lower}) as follows:
%%%%%%%%%%%%%%
\vspace{0.1cm}
\begin{corollary}\label{cor:dif} For any $i_k \notin \hat{\Ic}^{(k-1)}$, the lower-bound in (\ref{eq:lower}) can be rewritten as
\begin{align}
(\ref{eq:lower}) + C_2 &= (1-\lambda_k)\log\left(p_{i_k}/(1-p_{i_k})\right)\nonumber\\
&\;\;\;\;\;\;+ \frac{1}{\sigma^2}(\qv^{(k)})^{\trasp}\av_{i_k} -\frac{1}{2\sigma^2}\tau^{(k)}\|\av_{i_k}\|_2^2,\label{eq:obj_bin}
%&\; \eqdef \Phi\left(j |\hat{\Ic}^{(k-1)},p_j,\beta,\rv^{(k)},\Am\right),\label{eq:obj_alg}
\end{align} for some constant $C_2$ with respect to $j$ and 
\begin{align*}
\qv^{(k)} &= \beta(1-\lambda_k)\rv^{(k)}- \beta^2\lambda_k(1-\lambda_{k+1})\Am_{| \hat{\Ic}^{(k-1)}}\onev \\
\tau^{(k)} &= \beta^2(1- 3\lambda_k + 2\lambda_k\lambda_{k+1}).
%\sum_{i \notin \hat{\Ic}^{(k-1)}} \av_i\\
%\tau^{(k)} &= \beta^2(1- 3\lambda_k + 2\lambda_k\lambda_{k+1}).
\end{align*} 
\end{corollary}
\begin{IEEEproof} The proof is provided in Section~\ref{proof:cor1}.
\end{IEEEproof}
\vspace{0.1cm}
Note that the lower-bound in (\ref{eq:obj_bin}) is referred to as {\em B-MAP proxy}, which will be used as an alternative of OMP proxy \cite{Tropp2007} and MAP-ratio proxy \cite{LEE2008}. The proposed greedy algorithm, based on B-MAP proxy, is described in {\bf Algorithm 1}. In Section~\ref{sec:simulations}, it will be shown that the propose B-MAP proxy can provide a significant gain compared to the existing ones. 

%This metric is referred to as B-MAP proxy, compared with the existing correlation-magnitude proxy in OMP \cite{Tropp2007} and MAP-ratio proxy in MAP-OMP \cite{LEE2008}.

%%%%%%%%%%%%%%%%
\subsection{Proof of Theorem~\ref{thm:lower}}\label{proof:thm1}

This section provides the proof of  Theorem~\ref{thm:lower}. Without loss of generality, we focus on the $k$-th iteration by assuming that the indices chosen during the pervious $k-1$ iterations are belong to the true support (i.e., $\hat{\Ic}^{(k-1)}=\{\hat{i}_1,...,\hat{i}_{k-1}\} \subset \Sc(\xv)$. Following the notation in the statement of Theorem~\ref{thm:lower}, we also let $\Ic = \hat{\Ic}^{(k-1)}\cup\{i_k\}$.  
With these notations, we can have:
\begin{align}
&\log\PP\left(i_k \in \Sc(\xv)|\hat{\Ic}^{(k-1)}\subset \Sc(\xv), \yv,\Am\right) + D_1\nonumber\\
&=\log{\PP(\Ic \subset \Sc(\xv)|\yv,\Am)}=\log{\PP(s_{\hat{i}_1}=1,...,s_{i_k}=1|\yv,\Am)}\nonumber\\
&= \log{\sum_{\uv \in \Omega_{\Ic}} \PP(\sv=\uv|\yv,\Am)}= \log{\sum_{\uv\in \Omega_{\Ic}}\frac{{p_{\sv}(\uv)}{f}_{\yv|\sv}(\yv|\uv)}{{f}_{\yv}(\yv)}}\nonumber\\
&= \log{\left(\frac{|\Omega_{\Ic}|}{f_{\yv}(\yv)}\right)\frac{1}{|\Omega_{\Ic}|}\sum_{\uv \in \Omega_{\Ic}}{p_{\sv}(\uv)}{f}_{\yv|\sv}(\yv|\uv)},\label{eq:MAP}
\end{align} where  $p_{\sv}$ and $f_{\yv|\sv}$ denote the joint PMF and conditional PDF, respectively, $\Omega_\Ic$ is defined in (\ref{eq:omega_sub}), and the constant term $D_1$ is equal to $D_1 = \log{\PP(\hat{\Ic}^{(k-1)}\subset \Sc(\xv) | \yv, \Am)}$.
%\begin{equation}
%D_1 =\log{\PP\left(\hat{\Ic}^{(k-1)}\subset \Sc(\xv) | \yv, \Am\right)}.
%\end{equation} 
Before deriving the lower-bound on (\ref{eq:MAP}), we first provide the useful definition which will be used for this proof. 
%%%%%%%%%%%%%%%%
\vspace{0.1cm}
\begin{definition}\label{eq:def2} Given $\Ic= \hat{\Ic}\cup\{i_k\}$, we define a length-$N$ auxiliary random vector $\tv$ which can take values from $\Omega_{\Ic}$ uniformly. Namely, $\tv$ is a binary random vector with sparsity-level $K$. We can easily see that each component of $\tv$ follows a Bernoulli random variable such as
\begin{align}
&t_j \sim  {\rm Bern}\left(\lambda_k \right),  j \notin \Ic \mbox{ and } t_j \sim {\rm  Bern(1)} ,  j \in \Ic. \label{eq:random_U}
\end{align}  Here, for consistency, we used the notation of ${\rm Bern}(1)$ to indicate the constant value 1.\flushright\QED
\end{definition}

To avoid the confusion, we remark that in the rest of this section, $\sv$, $\xv$, $\yv$, and $\tv$ represent random vectors and $\uv$  is used to represent a constant vector. From Definition~\ref{eq:def2}, (\ref{eq:MAP}) can be rewritten as
\begin{align}
&\log \PP(\Ic\subset \Sc(\xv)|\yv,\Am) + D_2 \nonumber\\
&=\log\frac{1}{|\Omega_{\Ic}|}\sum_{\uv \in \Omega_{\Ic}}{p_{\sv} (\sv)}{f}_{\yv|\sv}(\yv|\sv)=\log \mathbb{E}_{\tv}\left[{p_{\sv}(\tv)}{f}_{\yv|\sv}(\yv|\tv)\right]\nonumber\\
&\stackrel{(a)}{\geq} \underbrace{\mathbb{E}_{\tv}\left[\log{p_{\sv}(\tv)}\right]}_{\mbox{A priori}}+\underbrace{\mathbb{E}_{\tv}\left[\log{{f}_{\yv|\sv}(\yv|\tv)}\right]}_{\mbox{Likelihood}}, \label{eq:MAP3}
\end{align} where (a) follows the Jensen's inequality due to the concavity of $\log$ function and $D_2 = \log{1/|\Omega_{\Ic}|}$.
 In the following subsubsections, we will derive the closed-form expressions for the above two parts, which will complete the proof.

%%%%%%%%%%%%%%%%%%%%%%%%%%%
\subsubsection{A priori part}
We derive a closed-form expression for a priori part in  (\ref{eq:MAP3}) by approximating the joint probability of $\sv$ such as
\begin{equation}
p_{\sv}(\uv) \approx \prod_{j=1}^N \PP(s_j = u_j)
\end{equation} for $u_j \in \{0,1\}$, where $\PP(s_j = 1) = p_j$. Then, we have:
\begin{align}
&\mathbb{E}_{\tv}\left[\log{p_{\sv}(\tv)}\right]= \mathbb{E}_{\tv}\left[\sum_{j=1}^{N}\log{\big(\mathbb{I}_{\{t_j=1\}}p_j + \mathbb{I}_{\{t_j =0\}}(1-p_j) \big)}\right]\nonumber\\
&=\sum_{j\in\Ic}\log{p_j}+\sum_{j\in [N]\setminus \Ic} \mathbb{E}\left[\log{(\mathbb{I}_{\{t_j=1\}}p_j + \mathbb{I}_{\{t_j =0\}}(1-p_j))}\right],\label{eq:apriori}
\end{align} where $\mathbb{I}_{\Ac}$ denotes an indicator function with $\mathbb{I}_{\Ac} = 1$ if an event $\Ac$ occurs, and $\mathbb{I}_{\Ac} = 0$, otherwise. Since $t_j$ is a Bernoulli random variable as in (\ref{eq:random_U}), we have:
\begin{align}
&\mathbb{E}_{t_j}[\log{(\mathbb{I}_{\{t_j=1\}}p_j + \mathbb{I}_{\{t_j =0\}}(1-p_j))}] \nonumber\\
&= \lambda_{k} \log{p_j} + \left(1-  \lambda_{k} \right) \log{(1-p_j)}\nonumber\\
&= -\Hc_2\left( {\rm Bern}(\lambda_{k}) \right) - \mathcal{D}_{{\rm KL}}\left({\rm Bern}(\lambda_{k}))\big|\big|{\rm Bern}(p_{j})\right),\label{eq:KL}
\end{align} where $\Hc_2$ and $ \mathcal{D}_{{\rm KL}}(\cdot || \cdot)$ denote the binary entropy function  and  KL divergence, respectively.
Plugging (\ref{eq:KL}) into (\ref{eq:apriori}), we have:
\begin{align}
&\mathbb{E}_{\tv}\left[\log{p_{\sv}(\tv)}\right] + D_3 \nonumber\\
&\;\;\;\;\;= \sum_{j\in\Ic}\log{p_j}- \sum_{j\in[N]\setminus \Ic} \mathcal{D}_{{\rm KL}}\left({\rm Bern}(\lambda_{k}))\big|\big|{\rm Bern}(p_{j})\right)\nonumber\\
&\;\;\;\;\;=\sum_{j\in [N]} - \mathcal{D}_{{\rm KL}}\left({\rm Bern}(\alpha_{j})\big|\big|{\rm Bern}(p_{j})\right),\label{eq:sim_apriori}
\end{align} where $\alpha_j = 1,  j \in \Ic$, $\alpha_j = \lambda_k, j \notin \Ic$, and  $D_3=\sum_{j\in [N]\setminus\Ic} \Hc_2\left( {\rm Bern}(\lambda_{k}) \right)$. This proves the first part (i.e., a priori part) of Theorem~\ref{thm:lower}.

%%%%%%%%%%%%%%%%%%%%%%%%%%%%%
\subsubsection{Likelihood part}
From the measurement model in  (\ref{PDF:channel}) and (\ref{eq:s_PDF}), we first have:
\begin{align}
&\mathbb{E}_{\tv}\left[\log{f}_{\yv|\sv}(\yv|\tv)\right] = \sum_{j=1}^{M}\mathbb{E}_{\tv}\left[\log{{f}_{y_j|\sv}({y}_{j}|\tv)}\right].\label{eq:part1}
%&= \sum_{j=1}^{M}\mathbb{E}_{\qv(\Omega_{\Ic})}\left[\log\left(\frac{1}{\sqrt{2\pi\sigma^2}}\exp\left({-\frac{{\left(y_{j}-V_j\right)}^2}{2\sigma^2}}\right)\right)\right].
\end{align} Here, each term in the summation is given as
\begin{align*}
&\mathbb{E}_{\tv}\left[\log{{f}_{y_j|\sv}({y}_{j}|\tv)}\right]=\log{\frac{1}{\sqrt{2\pi\sigma^2}}}- \mathbb{E}_{\tv}\left[(r^{(k)}_{j}-\beta v_j)^2/2\sigma^2\right],%\label{eq:part2}
\end{align*} where $\rv^{(k)} = \yv - \sum_{j\in \hat{\Ic}^{(k-1)}}  \beta \av_{j}$ and $\vv = \Am_{|\hat{\Ic}^{(k-1)}}\tv_{|\hat{\Ic}^{(k-1)}}$. Focusing  only on the interesting terms associated with $i_k$, we can obtain:
\begin{align}
&\mathbb{E}_{\tv}\left[\log{f}_{\yv|\sv}(\yv|\tv)\right] + D_4  \nonumber\\
%&=\frac{1}{\sigma^2}\left(\frac{\beta}{\sigma^2} \sum_{j=1}^{M} r^{(k)}_{j}\mathbb{E}_{\tv}[v_j]-\frac{\beta^2}{2\sigma^2}\mathbb{E}_{\tv}[v_j^2]\right)\nonumber\\
&\;\;\;\;\;\;\;=\frac{\beta}{\sigma^2}\left(\rv^{(k)}\right)^{\trasp}\EE_{\tv}[\vv]-\frac{\beta^2}{2\sigma^2}\trace\left(\EE[\vv\vv^{\trasp}]\right),
%&\;\;=\frac{1}{\sigma^2}\left(\rv^{(k)}\right)^{\trasp} \Am_{|\hat{\Ic}}\EE_{\tv}[\tv_{|\hat{\Ic}^{(k-1)}}]-\frac{1}{2\sigma^2}\trace\left((\Am_{|\hat{\Ic}})^{\trasp}\Am_{|\hat{\Ic}}\Rm^{(i_k)}\right),
\label{eq:likelihood}
\end{align} where  the constant term $D_4$ is given as
\begin{equation}
D_4 = - M\log{\frac{1}{\sqrt{2\pi\sigma^2}}} + \sum_{j=1}^{M} \frac{1}{2\sigma^2}\left(r^{(k)}_j\right)^2.
\end{equation}
From Definition~\ref{eq:def2}, we can easily compute the above expectations such as
%\beta\left(\av_{i_k} + \lambda_k \sum_{j \notin \Ic}\av_j\right)
\begin{align}
\EE_{\tv}[\vv] &=\beta \av_{i_k}+\sum_{j \notin \Ic}\beta \av_j \EE[t_j] =\beta\left(a_{i_k} + \lambda_k \Am_{|\Ic}\onev_{|\Ic}\right).\label{eq:l1}
% \sum_{j=1}^{N-k+1} \qv_j\EE[U_j]\nonumber\\
%&= \sum_{j=1}^{N-k+1} \qv_j \PP(U_{\pi_{k-1}^{-1}(j)} = 1)\EE[T_j|U_{\pi_{k-1}^{-1}(j)} = 1] \nonumber\\
%&=\hat{x}_{i_k} \qv_{\pi_{k-1}(i_k)} + \lambda_{k}\cdot m_X \sum_{j=1:j\neq \pi_{k-1}(i_k)}^{N-k+1} \qv_j, \label{eq:l1}
\end{align} %where we used the fact that $\bar{U}$ is the support random vector of $\bar{T}$. 
Similarly, from Definition~\ref{eq:def2}, we can get
\begin{align}
\trace\left(\EE[\vv\vv^{\trasp}]\right) & = \trace\left((\Am_{|\hat{\Ic}^{(t-1)}})^{\trasp}\Am_{|\hat{\Ic}^{(t-1)}}\Rm^{(i_k)}\right),
\end{align} where the $(i,j)$th component of $\Rm^{(i_k)}$ is obtained as
\begin{equation}
\Rm^{(i_k)}_{i,j} =  \begin{cases}
\beta^2, & i=j=\pi_{k-1}(i_k)\\
%\lambda_k,& i=j\neq \pi_{k-1}(i_k)\\
\beta^2\lambda_k\lambda_{k+1}, & i,j \notin\{\pi_{k-1}(i_k)\}, i\neq j\\
\beta^2 \lambda_k, & \mbox{ else}.
\end{cases}
\end{equation}
Finally, from (\ref{eq:sim_apriori}) and (\ref{eq:likelihood}), we obtain the following lower-bound:
\begin{align*}
&\log\PP(i_k \in \Sc(\xv)|\hat{\Ic}^{(k-1)}\subset \Sc(\xv), \yv, \Am) \geq (\ref{eq:sim_apriori}) + (\ref{eq:likelihood}) - C_1,
\end{align*} where the constant term with respect to $i_k$ is given as
\begin{align}
C_1 = \sum_{j=1}^{4}D_j. \label{eq:C1}
%&=\log{\frac{1}{|\Omega_{\Ic}|}\PP(\hat{\Ic}^{(k-1)}\subset \Sc(\xv) | \yv, \Am)} +  \sum_{j\in [N]\setminus\Ic} \Hc_2\left( {\rm Bern}(\lambda_{k}) \right)\nonumber\\
%&\;\; - M\log{\frac{1}{\sqrt{2\pi\sigma^2}}} + \sum_{j=1}^{M} \frac{1}{2\sigma^2}\left(r^{(k)}_j\right)^2\label{eq:C1}.
\end{align}
This completes the proof of Theorem~\ref{thm:lower}.

%%%%%%%%%%%%%%%%
\subsection{Proof of Corollary~\ref{cor:dif}}\label{proof:cor1}
For the ease of expression, we define the following positive semidefinite matrix:
\begin{equation}
\Bm\eqdef(\Am_{|\hat{\Ic}^{(k-1)}})^{\trasp}(\Am_{|\hat{\Ic}^{(k-1)}}). \label{eq:def}
\end{equation}  and also define the lower-bound in (\ref{eq:lower}) as $h_{\rm L}(j)$ for $j \notin \hat{\Ic}^{(k-1)}$. Then, for any $i,j \notin \hat{\Ic}^{(k-1)}$, we have the difference of lower-bounds as
\begin{align}
&h_{\rm L}(j) - h_{\rm L}(i) \nonumber\\
 &= (1-\lambda_k)\log\left(\frac{p_j}{1-p_j}\right)- (1-\lambda_k)\log\left(\frac{p_i}{1-p_i}\right)\nonumber\\
&+\frac{\beta}{\sigma^2}(\rv^{(k)})^{\trasp}\Am_{|\hat{\Ic}^{(k-1)}}\bv -\frac{\beta^2}{2\sigma^2} \trace(\Bm(\Rm^{(j)}-\Rm^{(i)})),\label{eq:ap}
\end{align} where $b_j = (\lambda_k - 1)$, $b_i = 1-\lambda_k$, and $b_\ell = 0$, for $\ell \in [N] \setminus\{i,j\}$. First, we obtain that
\begin{equation}
(\rv^{(k)})^{\trasp}\Am_{|\hat{\Ic}^{(k-1)}}\bv = (1-\lambda_k)(\rv^{(k)})^{\trasp}(\av_j- \av_i).\label{result1}
\end{equation}
We next consider the computation of the following term:
\begin{equation}
 \trace(\Bm(\Rm^{(j)}-\Rm^{(i)})) \eqdef \trace(\Rm).
\end{equation} It is noticeable that due to the similarity of $\Rm^{(j)}$ and $\Rm^{(i)}$, their difference $\Rm^{(j)}-\Rm^{(i)}$ is very sparse, i.e., it only has the non-zero components at the $i$-th column and rows, and $j$-th columns and rows. This sparsity makes it easy to compute the diagonal components of $\Rm$ (accordingly, $\trace(\Rm)$).  Letting  $\zeta_k = (\lambda_k-\lambda_k\lambda_{k+1})$, we obtain the diagonal components of $\Rm$ such as
\begin{align}
\Rm_{i,i} &= -\zeta_k \Bm(i,:)\onev + \zeta_k \Bm_{i,j}+ \left(\zeta_k + \lambda_k-1)\right)\Bm_{i,i}  \label{eq1}\\
\Rm_{j,j}&=\zeta_k \Bm(j,:)\onev - \zeta_k \Bm_{j,i} + \left(-\zeta_k + 1 - \lambda_k\right)\Bm_{j,j}\label{eq2}\\
\Rm_{\ell,\ell}&=-\zeta_k \Bm_{\ell,i} +\zeta_k \Bm_{\ell,j}, \mbox{ for } \ell\neq i,j. \label{eq3}
\end{align} 
From (\ref{eq1}), (\ref{eq2}), and (\ref{eq3}), we have:
\begin{align}
\sum_{\ell=1:\ell\neq i, j}^{N} \Rm_{\ell,\ell} &=-\zeta_k \onev^{\trasp}\Bm(:,i) + \zeta_k \onev^{\trasp}\Bm(:,j) \nonumber\\
&\;\;\;\;+\zeta_k (\Bm_{i,i} + \Bm_{j,i}) -\zeta_k(\Bm_{i,j}+\Bm_{j,j}). \label{eq4}
\end{align}  Since $\Bm$ is a symmetric matrix, we have that  $\Bm_{i,j} = \Bm_{j,i}$ and $((\Bm(j,:) - \Bm(i,:))\onev)^{\trasp} = \onev^{\trasp}(\Bm(:,j) - \Bm(:,i))$. Using this fact, we can get
\begin{align}
 %\trace\left(\Bm(\Rm^{(j)}-\Rm^{(i)})\right) 
 &\sum_{\ell=1}^{N} \Rm_{\ell,\ell} = (\ref{eq1}) + (\ref{eq2}) + (\ref{eq4}) \nonumber\\
&= 2\zeta_k \onev^{\trasp}(\Bm(:,j) - \Bm(:,i)) +(1 - 2\zeta_k - \lambda_k)(\Bm_{j,j} - \Bm_{i,i})\nonumber\\
&\stackrel{(a)}{=}2\zeta_k (\Am_{|\hat{\Ic}^{(k-1)}}\onev)^{\trasp}(\av_j - \av_i) +(1 - 2\zeta_k - \lambda_k)(\|\av_{j}\|_2^2 - \|\av_i\|_2^2),\label{result2}
\end{align} where (a) is due to the fact that $\Bm_{\ell,\ell} = \|\av_{\ell}\|_2^2$ and $\Bm(:,j) - \Bm(:,i) = (\Am_{|\hat{\Ic}^{(k-1)}})^{\trasp}(\av_j - \av_i)$. From  (\ref{result1}) and (\ref{result2}), we obtain
\begin{align}
&\frac{\beta}{\sigma^2}(\rv^{(k)})^{\trasp}\Am_{|\hat{\Ic}^{(k-1)}}\bv -\frac{\beta^2}{2\sigma^2} \trace(\Bm(\Rm^{(j)}-\Rm^{(i)}))=\nonumber\\
%&=\frac{\beta}{\sigma^2}(1-\lambda_k)(\rv^{(k)})^{\trasp}(\av_j- \av_i) - \frac{\beta^2}{\sigma^2}\zeta_k (\Am_{|\hat{\Ic}^{(k-1)}}\onev)^{\trasp}(\av_j - \av_i)-\frac{\beta^2}{2\sigma^2}(1 - 2\zeta_k - \lambda_k)(\|\av_{j}\|_2^2 - \|\av_i\|_2^2)\nonumber\\
&\frac{1}{\sigma^2}((\qv^{(k)})^{\trasp}\av_j -\frac{1}{2}\tau^{(k)}\|\av_j\|_2^2) - \frac{1}{\sigma^2}((\qv^{(k)})^{\trasp}\av_i -\frac{1}{2}\tau^{(k)}\|\av_i\|_2^2),\label{eq:LL}
\end{align}where $\qv^{(k)} = \beta(1-\lambda_k)\rv^{(k)} - \beta^2(\lambda_k-\lambda_k\lambda_{k+1})\Am_{| \hat{\Ic}^{(k-1)}}\onev$ and $\tau^{(k)} = \beta^2(1- 3\lambda_k + 2\lambda_k\lambda_{k+1})$.
%\begin{align*}
%\qv^{(k)} &= \beta(1-\lambda_k)\rv^{(k)} - \beta^2(\lambda_k-\lambda_k\lambda_{k+1})\Am_{| \hat{\Ic}^{(k-1)}}\onev\\
%\sum_{\ell \notin \hat{\Ic}^{(k-1)}} \av_\ell\\
%\tau^{(k)} &= \beta^2(1- 3\lambda_k + 2\lambda_k\lambda_{k+1}).
%\end{align*} 
Finally, from (\ref{eq:ap}) and (\ref{eq:LL}), we obtain the lower-bound, which only includes the terms associated with $j$, such as
\begin{align}
(\ref{eq:lower}) + C_2 &=(1-\lambda_k)\log(p_j/(1-p_j))\nonumber\\
&\;\;\;\;\;\;\;\;\;\; + \frac{1}{\sigma^2}((\qv^{(k)})^{\trasp}\av_j -\frac{1}{2}\tau^{(k)}\|\av_j\|_2^2),
\end{align} for some constant $C_2$. This completes the proof of Corollary~\ref{cor:dif}.

%%%%%%%%%%%%%%%%%%%%%%%%%%%%%%%%%%
\section{Asymptotic Analysis\\ for Exact Support Recover}\label{section:AAE}

In this section, we perform an asymptotic analysis for the exact support recover of the proposed greedy algorithm (see Algorithm 1).  It is assumed that  $\xv$ is a binary sparse signal (e.g., $\beta = 1$) as the impact of magnitude $\beta$ can be completely captured by the noise variance $\sigma^2$. For the analysis, we first derive a lower-bound on the success probability that a true support is successfully recovered by the proposed greedy algorithm. Based on this, we identify a scaling law on the required number of measurements for the perfect recovery especially when $N$ and $K$ go to infinity with  $\lim_{N\rightarrow \infty} \frac{K}{N} = \delta_K \mbox{ for some } 0< \delta_K <  1$. We first provide the technical lemma for the proof of our main result.
\vspace{0.1cm}
\begin{lemma}\label{tlem} Suppose that all the elements of $\av_i \in \RR^{M}, i \in [N]$ are drawn from IID Gaussian distribution with zero mean and variance $\frac{1}{M}$. Then,  the distribution of the scalar random variable $\frac{\av_{\ell}^{\trasp}(\av_i - \av_j)}{\|\av_i- \av_j\|_2}$ for $\ell\neq i,  j$, is Gaussian with zero mean and variance $\frac{1}{M}$, i.e., $\frac{\av_{\ell}^{\trasp}(\av_i - \av_j)}{\|\av_i-\av_j\|_2} \sim \Nc(0,\frac{1}{M})$.
\end{lemma}
\begin{IEEEproof} Let $\vv = \frac{\av_i - \av_j}{\|\av_i-\av_j\|_2}$. Then, we have that
\begin{align}
\vv^{\trasp}\vv &= \frac{(\av_{i}-\av_{j})^{\trasp}(\av_i - \av_j)}{\|\av_i-\av_j\|_2^2}=1,
\end{align}which implies that $\vv^{\trasp}\vv =1$ with probability 1. Then,  for some constant vector $\uv$, we can obtain the following conditional expectation and variance:
\begin{align}
\EE\left[\av_{\ell}^{\trasp}\vv|\vv=\uv\right]&=0\label{eq:mean}\\
\mbox{Var}\left[\av_{\ell}^{\trasp}\vv|\vv=\uv\right] &=\EE\left[\uv^{\trasp}\av_{\ell}\av_{\ell}^{\trasp}\uv\right]\stackrel{(a)}{=}\frac{1}{M}\uv^{\trasp}\uv = \frac{1}{M},\label{eq:variance}
\end{align} where (a) is due to the fact that the realization of $\vv^{\trasp}\vv$ is equal to 1 with probability 1. Using (\ref{eq:mean}) and (\ref{eq:variance}), we have:
\begin{align*}
f_{\av_{\ell}^{\trasp}\vv}(x)&=\int_{\uv} f_{\av_{\ell}^{\trasp}\vv | \vv}(x|\uv)f_{\vv}(\uv)d\uv\\
&=\int_{\uv} (1/\sqrt{2\pi(1/M)})e^{-\frac{x^2}{2(1/M)}}f_{\vv}(\uv)d\uv\\
&=1/(\sqrt{2\pi(1/M)})e^{-\frac{x^2}{2(1/M)}},
\end{align*}which implies that $\av_{\ell}^{\trasp}\vv$ is a Gaussian random variable with zero mean and variance $1/M$. This completes the proof.
\end{IEEEproof}

%%%%%%%%%%%%%%%%%%%%%%%%%%%%%%%%
\begin{definition} We define the signal-to-noise ratio ($\SNR$) by
\begin{equation}
\mbox{\SNR} = \EE[\|\xv\|_2^2]/\EE[\|\zv\|^2] = K/(M\sigma^2). 
\end{equation} For the asymptotic analysis, it is assumed that 
\begin{equation}
\lim_{M,K\rightarrow \infty} \SNR = \delta_{\SNR},\label{eq:d_snr}
\end{equation} for some strictly positive $\delta_{\SNR}>0$. Namely, the power of a sparse signal is not disappeared compared to the noise power in the large system limit. \;\;\;\;\;\;\;\;\;\;\;\;\;\;\;\;\;\;\;\;\;\;\;\;\;\;\;\;\;\;\;\;\;\;\;\;\;\;\;\;\;\;\;\;\;\;\;\;\;\;\;\;\QED
\end{definition}

The following theorem states the scaling law on the required measurements.
\vspace{0.1cm}
\begin{theorem} Suppose that all the components of a measurement matrix $\Am$ follow i.i.d. Gaussian distributions with zero mean and variance $\frac{1}{M}$. Then, the proposed greedy algorithm can perfectly recover a $K$-sparse binary signal with $M$ noisy measurements within $K$ number of iterations, provided that the number of measurements scales as
\begin{equation}
M=\Oc\left((1+1/\delta_{\SNR})K\ln(N)\right),
\end{equation} where $N$ and $K$ go to infinity with $\lim_{N\rightarrow\infty} \frac{K}{N} = \delta_K$ for some strictly positive $\delta_K$. 
\end{theorem}
\begin{IEEEproof}
Assuming that $\hat{\Ic}^{(k-1)} \subset \Sc$ (i.e., $(k-1)$ indices are successfully recovered), we focus on the $k$-th iteration of the proposed greedy algorithm. For any two indices  $i \in \Sc(\xv) \setminus \hat{\Ic}^{(k-1)}$ and $ j \notin \Sc(\xv)$, the corresponding B-MAP proxies are given as
\begin{align}
\gamma_i^{(k)} &=(\qv^{(k)})^{\trasp}\av_i -\frac{1}{2}\tau^{(k)}\|\av_i\|_2^2\nonumber\\
&=\frac{1}{2}(1-\lambda_k)\|\av_i\|_2^2 + (1-\lambda_k)\zv^T\av_i  \nonumber\\
&\;\;\; + (1-\lambda_k)\sum_{\ell \in \Sc(\xv)\setminus \hat{\Ic}^{(k-1)}\cup\{i\}}\av_{\ell}^{\trasp}\av_i \nonumber\\
&\;\;\; -\lambda_k(1-\lambda_{k+1})\sum_{\ell \in [N]\setminus  \hat{\Ic}^{(k-1)}\cup\{i,j\} }\av_{\ell}^{\trasp}\av_i \nonumber\\
&\;\;\; -\lambda_k(1-\lambda_{k+1})\av_{i}^{\trasp}\av_j,\label{gamma1}\\
\gamma_j^{(k)} &=(\qv^{(k)})^{\trasp}\av_j -\frac{1}{2}\tau^{(k)}\|\av_j\|_2^2\nonumber\\
&=-\frac{1}{2}(1-\lambda_k)\|\av_j\|_2^2 + (1-\lambda_k)\zv^T\av_j \nonumber\\
&\;\;\; + (1-\lambda_k)\sum_{\ell \in \Sc(\xv)\setminus \hat{\Ic}^{(k-1)}\cup\{i\} }\av_{\ell}^{\trasp}\av_j \nonumber\\
&\;\;\; -\lambda_k(1-\lambda_{k+1})\sum_{\ell \in [N]\setminus \hat{\Ic}^{(k-1)}\cup\{i,j\} }\av_{\ell}^{\trasp}\av_j\nonumber\\
&\;\;\;  -\lambda_k(1-\lambda_{k+1})\av_{i}^{\trasp}\av_j + (1-\lambda_k)\av_i^{\trasp}\av_{j}.\label{gamma2}
\end{align} We notice that $\gamma_i^{(k)}$ and $\gamma_j^{(k)}$ are random variables as functions of an additive noise $\zv$, a random measurement matrix $\Am$, and support $\Sc(\xv)$. Based on this, we will derive the following error probability:
\begin{align}
\PP(\gamma_i \leq \gamma_j) &= \PP\left((\gamma_i - \gamma_j)/\|\av_i-\av_j\|_2 \leq  0 \right).
\end{align} 
From (\ref{gamma1}) and (\ref{gamma2}), we have:
\begin{align}
\frac{\gamma_i - \gamma_j}{\|\av_i-\av_j\|_2} &=\frac{1}{2}(1-\lambda_k)\frac{\|\av_i\|_2^2 + \|\av_j\|_2^2-2\av_{i}^{\trasp}\av_j}{\|\av_i-\av_j\|_2} + \tilde{z}_{i,j}\nonumber\\
&=\frac{1}{2}(1-\lambda_k)\|\av_i - \av_j\|_2+ \tilde{z}_{i,j},\label{eq:mag}
\end{align} where the so-called effective noise $\tilde{z}_{i,j}$ is given as
\begin{align*}
\tilde{z}_{i,j} &=(1-\lambda_k)\zv^{\trasp}\vv +(1-2\lambda_k+\lambda_k\lambda_{k+1}) \sum_{\ell \in \Sc\setminus \Ic^{(k-1)}\cup\{i\}}\av_{\ell}^{\trasp}\vv\nonumber\\
&-\lambda_k(1-\lambda_{k+1}) \sum_{\ell \in [N]\setminus  \Sc \cup\{j\} }\av_{\ell}^{\trasp}\vv,
\end{align*} where $\vv= \frac{\av_i - \av_j}{\|\av_i-\av_j\|_2}$.
Clearly, we have $\EE[\tilde{z}_{i,j}] = 0$ and using the fact that $\av_{\ell}^{\trasp}\vv \sim \Nc(0,1/M)$ from Lemma~\ref{tlem}, 
we have:
\begin{align}
\sigma_{\tilde{z}_{i,j}}^2
&=\left(\frac{N-K}{N-k}\right)^2\frac{1}{M}\left(M\sigma^2+\left((K-k)\frac{N-K-1}{N-k-1}\right)\right).\label{eq:noi}
\end{align} From (\ref{eq:mag}) and (\ref{eq:noi}), the conditional error probability is obtained as 
\begin{align}
&\PP\left(\gamma_i - \gamma_j \leq  0 |\|\av_i\|_2, \|\av_j\|_2 \right)\nonumber\\
&\;\;\;\;\;\;\;\;\;\; = Q\left(\frac{\|\av_i - \av_j\|_2}{2\sqrt{\frac{1}{M}(M\sigma^2+(K-k)\frac{N-K-1}{N-k-1})}}\right) \nonumber\\
&\;\;\;\;\;\;\;\;\;\; \leq \exp\left(- \frac{M\|\av_i - \av_j\|_2^2}{8(M\sigma^2+(K-k)\frac{N-K-1}{N-k-1})}\right),
%&\approx\exp\left(- \frac{1}{2\left(\sigma^2+\frac{1}{M}\left(\left[1+(K-k)\frac{N-K-1}{N-k-1}\right]\right)\right)}\right)
\end{align} where $Q(\cdot)$ denotes the well-known Q-function and the last inequality follows the Chernoff bound of Q-function (i.e., $Q(x)\leq e^{-x^2/2}$). Finally, we have:
\begin{align}
&\PP\left(\gamma_i - \gamma_j \leq  0 \right)=\EE\left[\PP\left(\gamma_i - \gamma_j \leq  0 |\|\av_i\|_2, \|\av_j\|_2 \right)\right]\nonumber\\
%&\leq \EE\left[ \exp\left(- \frac{M\|\av_i - \av_j\|_2^2}{8\left(M\sigma^2+(K-k)\frac{N-K-1}{N-k-1}\right)}\right)\right]\nonumber\\
%&\stackrel{(a)}{\leq} \exp\left(- \frac{M\EE\left[\|\av_i - \av_j\|_2^2\right]}{8\left(M\sigma^2+(K-k)\frac{N-K-1}{N-k-1}\right)}\right)\nonumber\\
&\;\;\;\;\;\;\;\;\;\;\;\;\;\;\;\;\; \stackrel{(a)}{\leq} \exp\left(- \frac{M}{4(M\sigma^2+(K-k)\frac{N-K-1}{N-k-1})}\right),\label{eq:upperp}
\end{align}where (a) is due to the convexity of the exponential function and $\EE\left[\|\av_i - \av_j\|_2^2\right]=2$.

We are now ready to derive the success probability that the proposed greedy algorithm recovers the support perfectly. 
Let $\Ec_{k}$ denote the event that the $k$-th estimated support is belong to the true support set. Then, the success probability is computed as
\begin{equation}
P_s = \PP \left(\cap_{k=1}^{K} \Ec_{k}\right) =\prod_{k=1}^{K} \PP(\Ec_{k}|\Ec_{1},...,\Ec_{k-1}).\label{eq:suc_prob}
\end{equation}  The lower-bound on $\PP(\Ec_{k}|\Ec_{1},...,\Ec_{k-1})$ can be computed as
\begin{align}
&\PP(\Ec_{k}|\Ec_{1},...,\Ec_{k-1})=\PP(\max_{i \in \Sc(\xv)\setminus \hat{\Tc}^{(k-1)}}\gamma_i > \max_{j \in [N]\setminus \Sc(\xv)} \gamma_j )\nonumber\\
&\geq \PP(\gamma_i > \max_{j \in [N]\setminus \Sc(\xv)} \gamma_j) \mbox{ for some }i\in \Sc(\xv)\setminus \hat{\Tc}^{(k-1)} \nonumber\\
&=\prod_{j \in [N]\setminus \Sc(\xv)}\PP(\gamma_i > \gamma_j)=\prod_{j \in [N]\setminus \Sc(\xv)} (1 - \PP(\gamma_i \leq \gamma_j))\nonumber\\
&\stackrel{(a)}{\geq} \left(1-\exp\left(- \frac{M}{4(M\sigma^2+(K-k)\frac{N-K-1}{N-k-1})}\right)\right)^{N-K}, \label{eq:suc_u}
%&= 1- \PP\left(\gamma_i \leq \min_{j \in [N]\setminus \Sc(\xv)} \gamma_j \right)\\
%&\geq 1 - Q\left(\frac{P^2 - \mu(\Am)(1+4(K-k))}{\sqrt{2P^2\sigma^2}}\right),
\end{align} where (a) is from the upper-bound of the error-probability in (\ref{eq:upperp}). From (\ref{eq:suc_prob}) and (\ref{eq:suc_u}), we derive the lower-bound for the success probability as
\begin{align}
P_s &\geq \prod_{k=1}^{K} \left(1-\exp\left(- \frac{M}{4(M\sigma^2+(K-k)\frac{N-K-1}{N-k-1})}\right)\right)^{N-K}\nonumber\\
&\geq\left(1-\exp\left(- \frac{M}{4(M\sigma^2+(K-1))}  \right)\right)^{K(N-K)},
%&=\left(1-\exp\left(- \frac{M}{4\left(\tilde{\sigma}^2+K-1)\right)}\right)\right)^{K(N-K)},
\end{align} where we used the fact that $\frac{N-K-1}{N-k-1}<1$ and  $K-k \leq K-1$ for all $k$.
Letting
\begin{equation}
M = 4(M\sigma^2+K-1)\ln(K(N-K)),
\end{equation}  and $K=\delta_K N$ for some $0<\delta_K<1$, we can further simplify the lower bound as
%we can further simplify the lower-bound as
%\begin{align}
%P_s \geq \left( 1 - \frac{1}{K(N-K)} \right)^{K(N-K)}.
%\end{align} Letting $K=\delta_K N$ for some $0<\delta_K<1$,  we have:
\begin{align}
P_s \geq \left( 1 - \frac{1}{\delta_K(1-\delta_K)N^2} \right)^{\delta_K(1-\delta_K)N^2}.
\end{align} From this, it is shown that $\lim_{N \rightarrow \infty} P_s = 1$ as follows:
\begin{align}
&\lim_{N \rightarrow \infty} \ln{P_s} = \lim_{N\rightarrow \infty} \delta_K(1-\delta_K)N^2\ln{\left( 1 - \frac{1}{\delta_K(1-\delta_K)N^2} \right)} \nonumber\\
&\stackrel{(a)}{=}\lim_{N \rightarrow \infty}\frac{(2\delta_K(1-\delta_K)N)(\delta_K(1-\delta_K)N^2)}{(\delta_K(1-\delta_K)N^2-1)}\frac{2}{\delta_K(1-\delta_K)N^3}\nonumber\\
&=\lim_{N \rightarrow \infty}\frac{4\delta_K(1-\delta_K)}{\delta_K(1-\delta_K)N^2 - 1} = 0,\nonumber
\end{align} where (a) follows the L'Hospital's rule. %Accordingly, we conclude that $\lim_{N \rightarrow \infty} P_s = 1$.
From the facts that $N>M>2K$ (the condition for a unique sparse solution) and accordingly, $\ln(K(N-K))=\ln(N-K)+\ln(K)\leq 2\ln(N-K)$,
%\begin{align}
%\ln(K(N-K))&=\ln(N-K)+\ln(K)\leq 2\ln(N-K)\\
%\frac{1}{\epsilon}&=\frac{M\sigma^2 + (K-1)}{K\sigma^2 + (K-1)} \leq \frac{1}{\delta_K}.
%\end{align}
we can see that a $K$-sparse binary signal is perfectly recovered within $K$ number of iterations, provided that the number of measurements scales with $8(M\sigma^2+K-1)\ln(N-K)$.
%as, at least, 
%\begin{align*}
%M&= \frac{4}{\epsilon}((1+\sigma^2)K-1)\ln(N(N-K))\\
 %&\leq \frac{8}{\epsilon}((1+\sigma^2)K-1)\ln(N-K)
%\end{align*} 
Therefore, the scaling law of the required number of measurements becomes 
\begin{equation}
M=\Oc\left((1+1/\delta_{\SNR})K\ln(N)\right),
\end{equation} since in the large system limit, it is assumed that $M\sigma^2 \rightarrow K/ \delta_{\SNR}$.
This completes the proof.

%$M\geq 4(M\sigma^2+K-1)\ln(N-K) = 4(\tilde{\sigma}^2+K-1)\ln(N-K)$, where the equality is due to the fact that $M\sigma^2 \rightarrow \tilde{\sigma}^2$ for a sufficiently large $M$. Thus, the scaling law of the required number of measurements becomes $M=\Oc\left((\tilde{\sigma}^2+K)\ln(N)\right)$, which completes the proof.
 \end{IEEEproof}

%%%%%%%%%%%%%%%%%%%%%%%%
%%%%%%%%%%%%%%%%%%%%%%%
\section{Extensions of The Proposed B-MAP Proxy}\label{sec:extension}

We first extend the proposed B-MAP proxy for a more general sparse signal with random non-zero values and then combine the proposed method with some advanced greedy algorithms.

%%%%%%%%%%%%%%%%%%%%%%%%%%%%
\subsection{B-MAP proxy: Random non-zero values}

In Section~\ref{sec:B-MAP}, we have proposed a novel greedy algorithm based on B-MAP proxy, with the assumption of a constant-value sparse signal (e.g., binary sparse signal). In some applications, however, the values of non-zero components of a sparse signal can be an arbitrary value. In this section, we extend the proposed B-MAP proxy for the case that $x_i$ is drawn from a continuous PDF $f_{x_i}$. Also, it is assumed that $f_{x_i}$'s are known as a priori information.

For some fixed values $x_j=\beta_j$ for $j\in[N]$,  we can obtain the following B-MAP proxy from (\ref{eq:obj_bin}) by simply replacing the common constant  $\beta$ with the associated constant values $\beta_j$:
\begin{align}
\gamma_{j}^{(k)}&=(\qv^{(k)})^{\trasp}\av_j - \frac{1}{2}\tau^{(k)}\|\av_j\|_2^2 \nonumber\\
&= \beta_j(1-\lambda_k)(\rv^{(k)})^{\trasp}\av_j - \frac{1}{2}\beta_j^2(1-\lambda_k)\|\av_j\|_2^2 \nonumber\\
&- \beta_j\lambda_k(1-\lambda_{k+1}) \sum_{i \notin \hat{\Tc}^{(k-1)}\cup\{j\}} \beta_i \av_i^{\trasp} \av_j.\label{eq:obj_rand}
\end{align} The major difference from the previous constant-value case is that  the actual realizations of random variables $x_i$'s are not known. Suppose that  $x^{\star}_i$ denotes the actual realization of $x_i$ for $i\in\Sc(\xv)$ where $\Sc(\xv)$ represents the true support. Assuming that $\hat{\Tc}^{(k-1)} \subset \Sc$ and the corresponding values are completely recovered, the residual vector can be represented as
\begin{equation}
\rv^{(k)} = \sum_{i \in \Sc\setminus\hat{\Tc}^{(k-1)}} x^{\star}_i \av_i + \zv. \label{eq:true_residual}
\end{equation}  Plugging (\ref{eq:true_residual}) into (\ref{eq:obj_rand}),  we can define the metrics according to the following two difference cases:

{\em Case 1:} For $j \in \Sc(\xv)\setminus\hat{\Tc}^{(k-1)}$, B-MAP proxy in (\ref{eq:obj_rand}) can be represented as
\begin{align}
\gamma_{1,j}^{(k)} &= \frac{1}{2}(1-\lambda_k)(2\beta_j x^{\star}_j - \beta_j^2) \|\av_j\|_2^2 + \tilde{z}_{1,j},
\end{align}where $\tilde{z}_{1,j}$ includes all the rest of terms.

{\em Case 2:} For  $j \notin \Sc(\xv)\setminus\hat{\Tc}^{(k-1)}$, B-MAP proxy in (\ref{eq:obj_rand}) can be represented as
\begin{align}
\gamma_{0,j}^{(k)} &= -\frac{1}{2}(1-\lambda_k)\beta_j^2 \|\av_j\|_2^2 + \tilde{z}_{0,j},
\end{align}where $\tilde{z}_{0,j}$ includes all the rest of terms.
%\begin{align*}
%\tilde{z}_{0,j} &= \beta_j(1-\lambda_k)\zv^{\trasp}\av_j + (1-\lambda_k)\beta_j\sum_{i \in \Sc\setminus\hat{\Tc}^{(k-1)}} \beta_j x^{\star}_i \av_i^{\trasp}\av_j-\lambda_k(1-\lambda_{k+1})\sum_{i\notin \hat{\Tc}^{(k-1)}\cup\{j\}}\beta_j\beta_i \av_{i}^{\trasp}\av_{j}.
%\end{align*}
For the special case of $ \beta_i = \beta$ for $i \in \Sc$ (i.e., a constant sparse signal), the above two metrics can be simplified as
\begin{align}
\gamma_{1,j}^{(k)} &= \frac{1}{2}(1-\lambda_k)\beta^2 \|\av_j\|_2^2 + \tilde{z}_{1,j}\\
\gamma_{0,j}^{(k)} &= -\frac{1}{2}(1-\lambda_k)\beta^2 \|\av_j\|_2^2 + \tilde{z}_{0,j},
\end{align} since  $(2\beta_j x^{\star}_j - \beta_j^2) = \beta^2$. This implies that $\gamma_{j}^{(k)}$ is positive with high-probability for $j \in \Sc(\xv)$, while it is negative with high probability for $ j \notin \Sc(\xv)$. This is the fundamental principle of the proposed B-MAP proxy for the case of a constant-value sparse signal. Going back to the case of a general sparse signal, we should carefully choose $\beta_j$ at least such that 
\begin{equation}
2\beta_j x^{\star}_j - \beta_j^2 > 0.\label{eq:cond0}
\end{equation} This is required to ensure the correctness of $\gamma_{j}^{(k)}$ with high probability. Since we do not know the actual realization $x^{\star}_j$, we can accomplish it in a probability sense. Specifically, $\beta_j$ should be chosen such that 
\begin{equation}
\PP\left(2\beta_j x_j > \beta_j^2\right) \geq 1 - \delta, \label{eq:cond1}
\end{equation} for some arbitrary small $\delta >0$. If $x_j$ is a continuous random variable which can take both positive and negative values, it is not available to find such constant $\beta_j$. In other words, we require at least two candidates with different signs. For this reason, we consider two different cases in the below.

We first assume that $x_j$ is the so-called {\em one-sided} random variable, i.e., it can take only either negative or positive value with high probability.  Without loss of generality, it is assumed that  $\PP(x_j>0)\geq 1-\delta_1$ for an arbitrary small $\delta_1 > 0$. In this case, we can satisfy the condition in (\ref{eq:cond1}) with a positive $\beta_j$ alone. Due to the identicalness of $x_j$, we can let $\beta_j = \beta$ for all $j \in [N]$. Thus, we can determine the $\beta^{\star}$ by taking the solution of 
\begin{align}
\beta^{\star} = \min\left\{m_X, \max\{\beta>0 :\PP(x_j \geq \beta/2) \geq 1-\delta \}\right\}.\label{eq:beta_s}
\end{align} In the above, we try to choose $\beta_j^{\star}$ as close to the mean as possible among all possible values that  satisfies the condition in (\ref{eq:cond1}).  Assuming $x_j$'s are i.i.d. random variables, B-MAP proxy can be defined as
\begin{equation}
\gamma_{j}^{(k)} (\beta^{\star})= (\qv^{(k)}(\beta^{\star}))^{\trasp}\av_j - \frac{1}{2}\tau^{(k)}(\beta^{\star})\|\av_j\|_2^2, \label{B_MAP_PROXY}
\end{equation}where $\beta^{\star}$ is given in (\ref{eq:beta_s}), and
\begin{align}
\qv^{(k)} (\beta^{\star})&= \beta^{\star}(1-\lambda_k)\rv^{(k)}\nonumber\\
&\;\;\;\;\;\;\;\;- (\beta^{\star})^2\lambda_k(1-\lambda_{k+1})\Am_{|\hat{\Ic}^{(k-1)}}\onev\label{q}\\
%\sum_{\ell \notin \hat{\Ic}^{(k-1)}} \av_\ell\label{q}\\
\tau^{(k)}(\beta^{\star}) &= (\beta^{\star})^2(1- 3\lambda_k + 2\lambda_k\lambda_{k+1}).\label{tau}
\end{align} In this case, using $\hat{\Ic}^{(k-1)}$ and B-MAP proxy, the $k$-th support index is determined as
%%%%%%%%%
\begin{equation}
\hat{i}_k = \argmax_{j\in [N]\setminus\hat{\Ic}^{(k-1)}} \gamma_{j}^{(k)} (\beta^{\star}).
% \Phi(j|\hat{\Ic}^{(k-1)}, p_j, \beta^{\star}, \rv^{(k)}, \Am),
\end{equation}
%where the objective function is defined as
%\begin{align}
%&\Phi(j|\hat{\Ic}^{(k-1)}, p_j, \beta^{\star}, \rv^{(k)}, \Am)\eqdef (1-\lambda_k)\log\left(p_j/(1-p_j)\right)\nonumber\\
%&\;\;\;\;\;\;\;\;\;\;\;\;\;\;\;\;\;\;\;\;\;\;\; + (\qv^{(k)}(\beta^{\star}))^{\trasp}\av_j - \frac{1}{2}\tau^{(k)}(\beta^{\star})\|\av_j\|_2^2,\label{eq:general_obj}
%\end{align} where $\qv^{(k)}(\beta^{\star})$ and $\tau^{(k)}(\beta^{\star})$ are defined in (\ref{q}) and (\ref{tau}), respectively.
%\vspace{0.2cm}
\begin{example} For example, $x_j \sim \mbox{unif}[a,b]$ for some non-negative $a<b$. In this case, we can choose %$beta^{\star} = \min\{2b - 2(1-\delta)(b-a), (a+b)/2\}$, 
\begin{equation}
\beta^{\star} = \min\{2b - 2(1-\delta)(b-a), (a+b)/2\},
\end{equation}
where we used $\delta = 0.001$ for simulations. We next consider a Gaussian random variable as  $x_j \sim \Nc(m_X,\sigma_X^2)$. From the fact that $Q(-3.0) \approx 0.999$, we obtain: 
%$\beta^{\star} = \min\{2\times(- 3.1\times \sigma_X +  m_X), m_X\}$
\begin{equation}
\beta^{\star} = \min\{2\times(- 3.1\times \sigma_X +  m_X), m_X\}
\end{equation} 
for $\delta = 0.001$. Since $\beta^{\star}$ should be positive, he variance should satisfy the $\sigma_{X} < \frac{m_x}{3.1}$ for a given mean  $m_X$.\flushright\QED
\end{example}

Next, we consider a more general case that  $x_j$ is an arbitrary continuous random variable. As mentioned before, it is not available to choose a constant value  $\beta_j$ that satisfies the condition in (\ref{eq:cond0}).
%\begin{equation}
%2\beta_j x^{\star}_j > \beta_j^2.\label{eq:cond2}
%\end{equation} 
This is because unknown realization $x^{\star}_j$ can be either positive or negative. We thus consider the following two hypothesis according to the sign of $x^{\star}_j$:
\begin{itemize}
\item For $x^{\star}_j > 0$, the condition holds by choosing
\begin{align*}
\beta_{+}^{\star} =&\max\{\beta>0:  \PP(x_j \geq \beta/2|x_j>0) \geq 1-\delta\}.
%&\mbox{ subject to }  \PP(x_j \geq \beta_j/2|x_j>0) \geq 1-\delta.
\end{align*} 
\item Likewise, for $x^{\star}_j < 0$, the condition holds by choosing
\begin{align*}
\beta_{-}^{\star} =&\max\{\beta<0:  \PP(x_j \leq \beta/2|x_j<0) \geq 1-\delta\}. %\arg\max_{\beta < 0} \;\; \beta\\
%&\mbox{ subject to } \PP(x_j \geq \beta_j/2|x_j < 0) \geq 1-\delta.
\end{align*}  
\end{itemize} Then, we choose $\beta^{\star} = \min\{|\beta^{\star}_{+}|,|\beta^{\star}_{-}|\}$, which can satisfy the 
condition in (\ref{eq:cond0}) simultaneously. With this, we can define B-MAP proxy as
\begin{equation}
\gamma_j^{(k)} (\beta^{\star})= \max\{\gamma_j^{(k)}(\beta^{\star}), \gamma_j^{(k)}(-\beta^{\star})\}. \label{PROXY_BOTH}
\end{equation}  As before, the proposed greedy algorithm is defined as
\begin{align}
\hat{i}_k = \argmax_{j\in [N]\setminus\hat{\Ic}^{(k-1)}}& \gamma_j^{(k)} (\beta^{\star}).
% \max\Big\{\Phi(j|\hat{\Ic}^{(k-1)}, p_j,\beta^{\star}, \rv^{(k)}, \Am), \nonumber\\
%&\;\;\;\;\;\;\;\;\; \Phi(j|\hat{\Ic}^{(k-1)}, p_j, -\beta^{\star}, \rv^{(k)}, \Am) \Big\},\label{eq:both}
\end{align}
%where recall $\Phi(\cdot)$ is defined in (\ref{eq:general_obj}).

%%%%%%%%%%%%%%%%%%
%          ALGORITHMS                  %
%%%%%%%%%%%%%%%%%%
\begin{algorithm}
\caption{B-MAP}
\begin{algorithmic}[1]
\State {\bf Input:} Measurement matrix $\Am \in \RR^{M\times N}$, noisy observation $\yv \in \RR^{M}$, signal value $\beta^{\star}$ (see (\ref{eq:beta_s})), noise-variance $\sigma^2$, sparsity-level $K$, and a priori probabilities $\{p_j: j\in [N]\}$.
\vspace{0.05cm}
\State {\bf Output:} Support $\hat{\Ic}^{(K)} = \{\hat{i}_1,...,\hat{i}_K\}$.
\vspace{0.05cm}
\State {\bf Initialization:} $\hat{\Ic}^{(0)} = \phi$ and $\rv^{(1)} = \yv$.% and $\Am_{|\hat{\Ic}^{(0)}} = \Am$.
\vspace{0.05cm}
\State {\bf Iteration:} $k=1:K$
\begin{itemize}
%\item Set $\Qm = \Am_{|\hat{\Ic}^{(k-1)}}$ and $\tv = \rv^{(k-1)}$.
%\item Set $i'$ as the minimum element of  $\hat{\Ic}^{(k-1)}$.
\item Update the B-MAP proxy $\gamma^{(k)}_j(\beta^{\star})$.

\item Select the largest index of B-MAP proxy:
\begin{align*}
     \hat{i}_k = \argmax_{j\in [N]\setminus\hat{\Ic}^{(k-1)}} \gamma^{(k)}_j(\beta^{\star}).
     %\Phi(j|\hat{\Ic}^{(k-1)}, p_j, \beta^{\star}, \rv^{(k)}, \Am),
\end{align*} 
%where $\Phi(\cdot)$ is defined in (\ref{eq:general_obj}) (or in  (\ref{PROXY_BOTH})
%\item {\bf Otherwise}, find the $k$-th  index $\hat{i}_k$ by taking the solution of
%\begin{align*}
%     \hat{i}_k = &\argmax_{j\in [N]\setminus\hat{\Ic}^{(k-1)}} \max\Big\{\Phi(j|\hat{\Ic}^{(k-1)}, p_j, \beta^{\star}, \rv^{(k)}, \Am), \\
%     &\;\;\;\;\;\;\;\;\;\;\;\;\;\;\;\;\;\;\;\;\;\;\; \Phi(j|\hat{\Ic}^{(k-1)}, p_j, -\beta^{\star}, \rv^{(k)}, \Am)\Big\}.
%\end{align*} %where $\Phi(\cdot)$ is defined in (\ref{eq:general_obj}).
\item Merge the support: $\hat{\Ic}^{(k)} = \hat{\Ic}^{(k-1)} \cup \{\hat{i}_k\}$.
%\begin{equation*}
 %\hat{\Ic}^{(k)} = \hat{\Ic}^{(k-1)} \cup \{\hat{i}_k\}.
 %\end{equation*}
\item Least-square: $\xv_{\hat{\Ic}^{(k)}}:=\argmin_{\xv} \big\|\yv-\Am_{\hat{\Ic}^{(k)}}\xv \big\|_2$.
%\begin{equation*}
%\xv_{\hat{\Ic}^{(k)}}:=\argmin_{\xv} \big\|\yv-\Am_{\hat{\Ic}^{(k)}}\xv \big\|_2.
%\end{equation*} 
\item Update the residual vector: $\rv^{(k+1)} = \yv - \Am_{\hat{\Ic}^{(k)}}\xv_{\hat{\Ic}^{(k)}}$.
%\begin{equation*}
%\rv^{(k+1)} = \yv -  \sum_{i\in\hat{\Ic}^{(k)}}\hat{x}_i\av_i.
%\end{equation*}
\end{itemize}
\end{algorithmic}
\vspace{0.1cm}
$\Diamond$ The B-MAP proxy is defined in (\ref{B_MAP_PROXY}) and (\ref{PROXY_BOTH}) for one-sided and both-sided sparse signals, respectively.

$\Diamond$ This algorithm encompasses Algorithm 1 with  $\beta^{\star} = \beta$ (i.e., the fixed signal value).
%$\Diamond$ One-sided sparse signal implies that each non-zero component of the sparse signal can have only positive (or negative) values with high probability.
\end{algorithm}
%%%%%%%%%%%%%%%%%%%%%%%%%%%%%%%

%%%%%%%%%%%%%%%%%%%%%%%%%%
\subsection{Advanced Greedy Algorithms}

We introduce advanced greedy algorithms based on the proposed B-MAP proxy. Specifically, we construct B-CoSaMP and B-SP by following the basic frameworks of CoSaMP \cite{Needell2010} and SP \cite{Dai2009}, respectively, having B-MAP proxy instead of OMP proxy. 

%%%%%%%%%%%%%%%%%%
%          ALGORITHMS                  %
%%%%%%%%%%%%%%%%%%
\begin{algorithm}
\caption{B-CoSaMP}\label{B-CoSaMP}
\begin{algorithmic}[1]
\State {\bf Input:} Measurement matrix $\Am \in \RR^{M\times N}$, noisy observation $\yv \in \RR^{M}$, signal value $\beta^{\star}$ (see (\ref{eq:beta_s})), noise-variance $\sigma^2$, sparsity-level $K$, and a priori probabilities $\{p_j: j\in [N]\}$.
\State {\bf Output:} Support $\hat{\Ic}^{(K)} = \{\hat{i}_1,...,\hat{i}_K\}$.

\State {\bf Initialization:} $\hat{\Ic}^{(0)} = \phi$ and $\rv^{(1)}= \yv$. 
\State {\bf Iteration:} Repeat $k$ until the stopping criterion is satisfied.
\begin{itemize}
\item Select the $K$ largest indices using B-MAP proxy:
\begin{align*}
       \hat{\Tc}^{(k)} = \argmax_{|{\Tc^{(k)}}|=K, j \in [N]} & \gamma_{j}^{(k)}(\beta^{\star}). 
      % \Phi(j|\hat{\Ic}^{(k-1)}, p_j,\beta^{\star}, \rv^{(k)}, \Am).
\end{align*} 
\item Merge support set:  ${\hat{\Ic}^{(k)}} = {\hat{\Ic}^{(k-1)}} \cup \hat{\Tc}^{(k)}$
\item Least-square: $\hat{\xv}^{(k)}:=\arg\min_{\xv} \|\yv-\Am_{\hat{\Ic}^{(k)}}\xv\|_2$
\item Prune K largest value: $\hat{\Ic}^{(k)} = \argmax_{|\hat{\Ic}^{(k)}|=K} \;\; {\hat{\xv}^{(k)}}$
\item Update the residual vector: $\rv^{(k)} = \yv - \Am_{\hat{\Ic}^{(k)}}\hat{\xv}_{\hat{\Ic}^{(k)}}^{(k)}$
\end{itemize}
\end{algorithmic}
\vspace{0.1cm}
$\Diamond$ The B-MAP proxy is defined in (\ref{B_MAP_PROXY}) and (\ref{PROXY_BOTH}) for one-sided and both-sided sparse signals, respectively.

%$\Diamond$ stopping criterion: either $\ev^{(t)} = \zerov$ or $t = t_{\rm max}$
%$\Diamond$ In the above, MAP detection can be changed into (\ref{eq:both}) according to the probability distribution of a sparse signal vector.
\end{algorithm}
%%%%%%%%%%%%%%%%%%%%%%%%%%%%%%%

%%%%%%%%%%%%%%%%%
\subsubsection{B-CoSaMP}

CoSaMP is an efficient iterative recovery algorithm which can ensure the same performance guarantees as the best optimization-based approach (e.g., $\ell_1$-minimization) with lower computational complexity \cite{Needell2010}. Differently from OMP, it identifies the  $L=2K$ largest indices  based on OMP proxy for each iteration 
among all the possible $N$ candidate indices. After merging the previously chosen indices, least-square (LS) estimation is performed to estimate a sparse signal and then, from which the support indices are updated. As usual, these procedures are repeated until a stopping criterion is satisfied. In \cite{LEE2008}, MAP-CoSaMP was proposed by replacing OMP proxy with MAP-ratio proxy. Leveraging the proposed B-MAP proxy in (\ref{B_MAP_PROXY}), we propose B-CoSaMP where the detailed algorithm is provided in {\bf Algorithm 3}. It is noticeable that differently from the existing ones in \cite{Needell2010,LEE2008}, B-CoSaMP chooses the $L=K$ instead of $L=2K$, which is numerically determined. Clearly, the proposed B-CoSaMP has the same computational complexity of $\Oc(MN)$ with both CoSaMP and MAP-CoSaMP.

%%%%%%%%%%%%%%
\subsubsection{B-SP}

In \cite{Dai2009}, another iterative recovery algorithm, named SP, was proposed. This method almost follows the basic structures of CoSaMP but it has the additional least-square estimation for each iteration, in order to improve the accuracy of an estimated sparse signal. As before, MAP-SP was proposed in  \cite{LEE2008}  by replacing proxy function. Likewise, we propose B-SP by using the proposed B-MAP proxy in (\ref{B_MAP_PROXY}). The detailed algorithm is described in {\bf Algorithm 4}. Note that all the SP-based methods choose the $L=K$ differently from CoSaMP-based methods. These methods have the same computational complexity as  $\Oc(MNK)$.

%%%%%%%%%%%%%%%%%%
%          ALGORITHMS                  %
%%%%%%%%%%%%%%%%%%
\begin{algorithm}
\caption{B-SP}\label{B-SP}
\begin{algorithmic}[1]
\State {\bf Input:} Measurement matrix $\Am \in \RR^{M\times N}$, noisy observation $\yv \in \RR^{M}$, signal value $\beta^{\star}$ (see (\ref{eq:beta_s})), noise-variance $\sigma^2$, sparsity-level $K$, and a priori probabilities $\{p_j: j\in [N]\}$.
\State {\bf Output:} Support $\hat{\Ic}^{(K)} = \{\hat{i}_1,...,\hat{i}_K\}$.

\State {\bf Initialization:} $\hat{\Ic}^{(0)} = \phi$ and $\rv^{(1)}= \yv$ 
\State {\bf Iteration:} Repeat $k$ until the stopping criterion is satisfied.
\begin{itemize}
\item Select the $K$ largest indices using B-MAP proxy:
\begin{align*}
       \hat{\Wc}^{(k)} = \argmax_{|\Wc_{j}^{(k)}|=K,j \in [N]}\; \gamma_{j}^{(k)}(\beta^{\star}).
       %& \Phi(j|\hat{\Ic}^{(k-1)}, p_j,\beta^{\star}, \rv^{(k)}, \Am).
\end{align*} 
\item Merge support set:  ${\hat{\Ic}^{(k)}} = {\hat{\Ic}^{(k-1)}} \cup \hat{\Wc}^{(k)}$.
\item Least-square: $\hat{\bv}^{(k)}:=\arg\min_{\bv} \|\yv-\Am_{\hat{\Ic}^{(k)}}\bv\|_2$.
\item Prune K largest value: $\hat{\Ic}^{(k)} = \argmax_{|\hat{\Ic}^{(k)}|=K} \; {\hat{\bv}^{(k)}}$.
\item Least-square:
$\hat{\xv}^{(k)}:=\arg\min_{\xv} \|\yv-\Am_{\hat{\Ic}^{(k)}}\xv\|_2$.
\item Update the residual vector :\;\;$\rv^{(k)} = \yv - \Am_{\hat{\Ic}^{(k)}}\hat{\xv}_{\hat{\Ic}^{(k)}}^{(k)}$.
\end{itemize}
\end{algorithmic}
\vspace{0.1cm}
$\Diamond$ The B-MAP proxy is defined in (\ref{B_MAP_PROXY}) and (\ref{PROXY_BOTH}) for one-sided and both-sided sparse signals, respectively.

%$\Diamond$ stopping criterion: either $\ev^{(t)} = \zerov$ or $t = t_{\rm max}$
%$\Diamond$ In the above, MAP detection can be changed into (\ref{eq:both}) according to the probability distribution of a sparse signal vector.
\end{algorithm}
%%%%%%%%%%%%%%%%%%%%%%%%%%%%%%%

%%%%%%%%%%%%%%%%%%%%%%%%%
\section{Numerical Results}\label{sec:simulations}

In this section, we provide the simulation results to demonstrate the superiorities of the proposed (advanced) greedy algorithms. The construction probability is used as performance metric and OMP and MAP-OMP based algorithms are used as benchmark methods. If not mentioned explicitly, a priori probability is assumed to be uniform, i.e., there is no a priori knowledge. Note that in this case, the a priori part (i.e., the first term in (\ref{B_MAP_PROXY}) is ignored and the likelihood part (i.e., the second term in (\ref{B_MAP_PROXY}) is only concerned.  For simulations, we consider various types of measurement matrices as
\begin{itemize}
\item {\bf Gaussian matrix}: Each component is drawn i.i.d. Gaussian random variable with zero-mean and variance $1/M$.
\item {\bf Uniform matrix I}: Each component is drawn i.i.d. uniform random variable over $[0,1]$.
\item {\bf Uniform matrix II}: Each component is drawn i.i.d. uniform random variable over $[-0.5,0.5]$.
\item {\bf Bernoulli matrix}: Each component is drawn i.i.d. Bernoulli random variable having either $0$ or $1$ with equal probability. 
\end{itemize}

\begin{figure}
\centerline{\includegraphics[width=9cm]{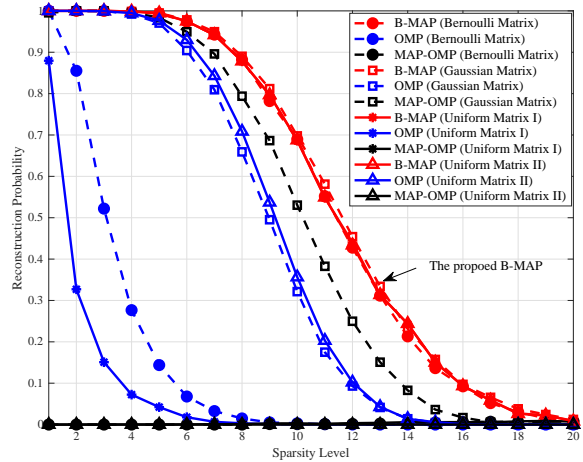}}
\caption{Performance comparisons of various greedy algorithms for the perfect reconstruction probability of a binary sparse signal with noise-free measurements. Here, we used $N=512$ and $M=64$.}
\label{sim:fig1}
\end{figure}

\begin{figure}
\centerline{\includegraphics[width=9cm]{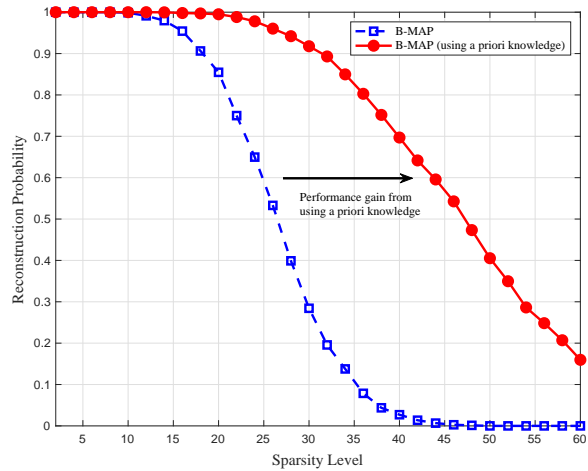}}
\caption{Performance improvement of the proposed greedy algorithm with a priori probability for the perfect reconstruction probability of a binary sparse signal with noisy measurements. Here, we used $N=256$, $M=128$, and $\SNR = 30$dB.}
\label{sim:fig2}
\end{figure}

{\bf Binary sparse signal:}  We examine the proposed B-MAP detection (in Algorithm 1), assuming a binary sparse signal. Here, each component of $\xv$ can be a support index with probability $0.5$ under the constraint of the sparsity-level $K$.  Fig.~\ref{sim:fig1} shows the reconstruction probabilities of B-MAP, OMP, and MAP-OMP for the above four types of measurement matrices. Since MAP-OMP is constructed with the assumption of Gaussian matrix above \cite{LEE2008}, it performs poor with the other measurement matrices. Even for Gaussian matrix, the proposed method can outperform MAP-OMP as well as OMP. More importantly, the proposed method can yield more uniform performance guarantees for various measurement matrices. In Fig.~\ref{sim:fig2}, we we verify that the proposed method can indeed improve the reconstruction performance by exploiting a priori information, while the other methods cannot use it. In this simulation, it is assumed that some of $K$ components have a priori probability of $0.55$ and the others have $0.5$. Gaussian matrix is also assumed. This result shows that if acquiring a priori knowledge on a sparse signal, the proposed can significantly improve the reconstruction probability by exploiting it appropriately.

%%%%%%%%%%%%%%%%%%%%%%%%%%%

%%%%%%%%%%%%%%%%%%%%%%%%%%%

\begin{figure}
\centerline{\includegraphics[width=9cm]{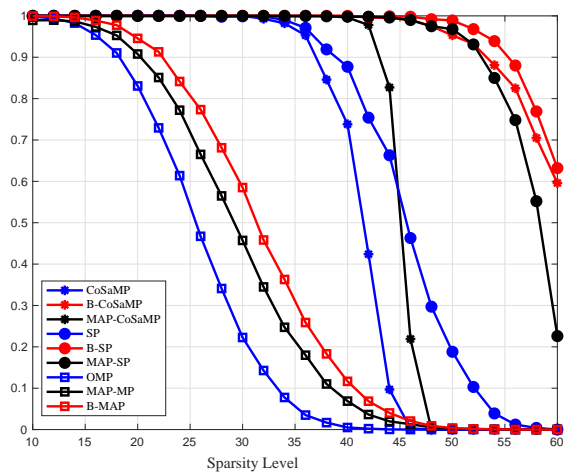}}
\caption{{\bf Gaussian matrix.} Performance comparisons of various (advanced) greedy algorithms for the perfect reconstruction probability of a sparse signal whose non-zero component is uniformly distributed between 0.5 and 1.5, with noisy measurements. Here, we used $N=256$, $M=128$, and $\SNR=30$dB.}
\label{sim:fig3}
\end{figure}

\begin{figure}
\centerline{\includegraphics[width=9cm]{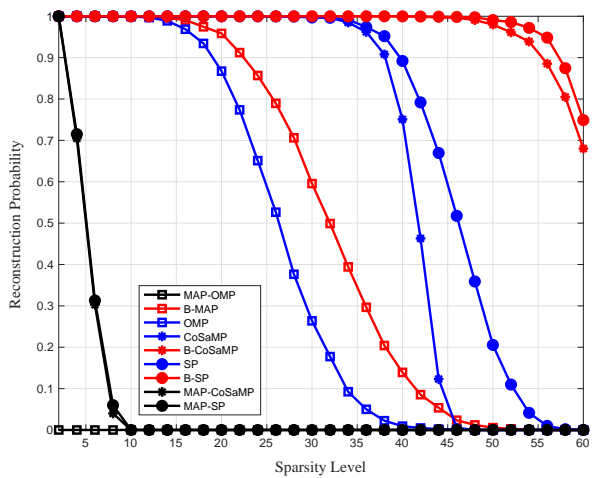}}
\caption{{\bf Uniform matrix II.} Performance comparisons of various (advanced) greedy algorithms for the perfect reconstruction probability of a sparse signal whose non-zero component is uniformly distributed between 0.5 and 1.5, with noisy measurements. Here, we used $N=256$, $M=128$, and $\SNR=30$dB.}
\label{sim:fig4}
\end{figure}

%%%%%%%%%%%%%%%
{\bf Random sparse signal:} We now consider the performances of the proposed (advanced) greedy algorithms for a more practical scenario in which non-zero components of a sparse signal are not a constant (e.g., binary) but are uniformly distributed between 0.5 and 1.5 (i.e., $x_i \sim {\rm Unif}[0.5,1.5]$). The corresponding reconstruction probabilities are provided in Figs.~\ref{sim:fig3} and~\ref{sim:fig4} for Gaussian and uniform matrices, respectively. As in the binary sparse signal, the proposed B-MAP detection yields a significant gain compared with the existing methods as
OMP and MAP-OMP. It seems that OMP can provide a good performance only when the components of a measurement matrix can both positive and negative (e.g., Gaussian and uniform II matrices). Regarding the advanced algorithms, 
 the proposed B-CoSaMP and B-SP achieve higher reconstruction probabilities than their counterparts  as CoSaMP and MAP-CoSaMP, and SP and MAP-SP, respectively. This verifies that the proposed B-MAP proxy is better to identify support indices than the correlation-magnitude proxy and MAP-ratio proxy. If not included in the paper, we confirmed that for the other measurement matrices, the proposed greedy algorithms yield a better performance than the corresponding counterparts and the gap is usually higher than that of Gaussian matrix.  From the above simulations, we can see that the proposed B-MAP proxy can be better alternative of the existing correlation-magnitude proxy in \cite{Tropp2007} and MAP-ratio proxy \cite{LEE2008}. Therefore, we expect that the proposed proxy can enhance the other advanced greedy algorithms as generalized OMP \cite{Wang2012} and multipath matching pursuit \cite{Kwon2014}.

%%%%%%%%%%%%%%%%%%%%%%%%%%%

%%%%%%%%%%%%%%%%%%%%%%%%%%%

%%%%%%%%%%%%%%%%%%%%%
\section{Conclusion}\label{sec:conclusion}

We proposed a novel greedy algorithm based on {\em bit-wise maximum a posteriori} (B-MAP) detection for a sparse recovery problem. This method performs an optimal detection for each greedy iteration, provided that all the detected indices in the previous iterations are correct. Furthermore, we considerably reduced the complexity of B-MAP detection by deriving a good proxy (named B-MAP proxy) on the optimization metric. It is remarkable that the proposed B-MAP proxy has the same computational complexity with the popular OMP proxy, only requiring vector correlations. We proposed advanced greedy algorithms, referred to as B-CoSaMP and B-SP, by simply replacing OMP-proxy with B-MAP proxy. Via simulations, we demonstrated that the proposed greedy algorithms yielded non-trivial gains compared with the counterpart methods based on OMP proxy and MAP-ratio proxy. Therefore, the proposed B-MAP proxy would a better alternative of the existing proxies. One interesting future work is to analyze the proposed advanced greedy algorithms asymptotically, in order to derive their scaling-laws. Another interesting future work is to develop novel advanced greedy algorithms by leveraging the idea of advanced tree-search algorithms as list, stack, and Fano decodings.

%Our ongoing work is to extend the proposed method for the case of general sparse signal vector with a certain probability distribution. Another interesting research direction is to consider sparse support recovery problems with multiple or quantized measurements. 

%%%%%% Reference %%%%%%

\end{document}